\documentclass[prb,aps,epsf,twocolumn,showpacs,10pt]{revtex4-1}
\usepackage{graphicx,amsfonts,times,bm,amsmath,verbatim,color,array}

\newcommand{\ua}{\uparrow}
\newcommand{\da}{\downarrow}
\newcommand{\ra}{\rightarrow}

\renewcommand{\>}{\rangle}
\renewcommand{\(}{\left(}
\renewcommand{\)}{\right)}

\begin{document}

\title{Majorana Fermions in Chiral Topological Ferromagnetic Nanowires}

\author{Eugene Dumitrescu$^1$}
\author{Brenden Roberts$^1$}
\author{Sumanta Tewari$^1$}
\author{Jay D. Sau$^2$}
\author{S. Das Sarma$^2$}

\affiliation{ $^1$Department of Physics and Astronomy, Clemson University, Clemson, SC 29634\\
$^2$ Condensed Matter Theory Center, Department of Physics, University of Maryland, College Park, MD 20742}

\begin{abstract}
Motivated by a recent experiment in which zero-bias peaks have been observed in scanning tunneling microscopy (STM) experiments performed on chains of magnetic atoms on a superconductor, we show, by generalizing earlier work, that a multichannel ferromagnetic wire deposited on a spin-orbit coupled superconducting substrate can realize a non-trivial chiral topological superconducting state with Majorana bound states localized at the wire ends.
The non-trivial topological state occurs for generic parameters requiring no fine tuning, at least for very large exchange spin splitting in the wire.
We theoretically obtain the signatures which appear in the presence of an arbitrary number of Majorana modes in multi-wire systems incorporating the role of finite temperature, finite potential barrier at the STM tip, and finite wire length.
These signatures are presented in terms of spatial profiles of STM differential conductance which clearly reveal zero energy Majorana end modes and the prediction of a multiple Majorana based fractional Josephson effect.
A substantial part of this work is devoted to a detailed critical comparison between our theory and the recent STM experiment claiming the observation of Majorana fermions in ferromagnetic atomic chains on a superconductor. 
The conclusion of this detailed comparison is that although the experimental observations are not manifestly inconsistent with our theoretical findings, the very small topological superconducting gap and the very high temperature of the experiment make it impossible to decisively verify the existence of a localized Majorana zero mode, as the spectral weight of the Majorana mode is necessarily spread over a very broad energy regime exceeding the size of the gap. 
Such an extremely broad (and extremely weak) conductance peak could easily arise from any sub-gap states existing in the rather complex system studied experimentally and may or may not have anything to do with a putative Majorana zero mode as discussed in the first half of our paper.
Thus, although the experimental findings are indeed consistent with a highly broadened and weakened Majorana zero bias peak, much lower experimental temperatures (and/or much larger experimental topological superconducting gaps) are necessary for any definitive conclusion.
\end{abstract}

\pacs{}
\maketitle

\section{Introduction}

In contrast to ordinary charged Dirac fermions (e.g. electrons, positrons) which come in oppositely charged particle anti-particle pairs, Majorana fermions are their own anti-particles \cite{Majorana,Wilczek}.
Although Majorana fermions, which are a special kind of neutral Dirac fermions, were first predicted in the context of high energy physics \cite{Majorana} (specifically, as a hypothetical model for neutrinos, which may or may not be Majorana fermions), a condensed matter analog of them has recently been proposed to exist as localized zero-energy quasiparticles bound to order parameter defects in topologically ordered systems\cite{Read-Green,Kitaev-1D}.
In low dimensional systems ($d\leq2$), these localized defect-bound zero-energy Majorana quasiparticles obey exotic non-Abelian quantum statistics \cite{Read-Green,Kitaev-1D,Ivanov} (and are therefore not any kind of Dirac fermions at all).
Due to their non-Abelian braiding statistics and non-local topological nature, these zero-energy Majorana bound states (MBS) can be used as the building blocks of a fault-tolerant quantum computer \cite{Kitaev-1D,Nayak_2008}.
Recently, realistic materials such as chiral $p$-wave superconductors e.g., strontium ruthenate\cite{Tewari-strontium}, topological insulator-superconductor interfaces \cite{Fu-Kane}, fermionic cold atom gases \cite{Zhang-Tewari,Sato-Fujimoto}, spin-orbit coupled semiconducting thin films \cite{Sau-Generic,Tewari-Annals,Long-PRB} and nanowires \cite{Long-PRB,Roman,Oreg} in proximity to conventional superconductors, all realizing an analog of topological spinless $p$-wave superconductors\cite{Read-Green,Kitaev-1D}, have been proposed as hosts for MBS.
Recent experimental observations in semiconductor-superconductor heterostructures \cite{Mourik,Deng,Weizman,Rokhinson,Churchill,Finck} seem to support the presence of MBS in condensed matter systems, although any conclusive evidence for non-Abelian statistics is still lacking \cite{Stanescu}.

In concurrent formal theoretical developments, recent work \cite{AZ, Schnyder_2008,Kitaev_2009,Ryu_2010} established that the quadratic Hamiltonians describing gapped topological insulators and topological superconductors (TS) can be classified into ten distinct topological symmetry classes, and that each is characterized by a topological invariant counting the number of topologically protected edge modes.
According to this classification, the experimentally investigated semiconductor-superconductor heterostructures \cite{Sau-Generic,Tewari-Annals,Long-PRB,Roman,Oreg,Mourik,Deng,Weizman,Rokhinson,Churchill,Finck} belong to the topological class D in which MBS are protected by the superconducting particle-hole (PH) symmetry.
Additionally, one-dimensional topological superconductors belonging to the time reversal class DIII\cite{Sato_11,Law_12,Deng_12,Keselman_13,Zhang_13,Deng_13,Law_14,Flensberg_13,Nakosai_13,Dumitrescu_TR} and BDI \cite{TS_BDI,minigap, Chakravarty,Diez,Wong_13,He} have recently been proposed.
DIII topological superconductivity is indexed by a ${\mathbb{Z}}_2$ topological invariant indicating the presence or absence of a Majorana Kramers pair, while in BDI topological superconductivity, for instance as recently proposed\cite{Dumitrescu} in the context of putative spin-triplet ferromagnetic superconductors such as the organic superconductors and lithium purple bronze ($\text{Li}_{0.9} \text{Mo}_{6} \text{O}_{17}$), a ${\mathbb{Z}}$ invariant counts the number of MBS.

Very recent experimental work \cite{Yazdani_14} suggests that atomic scale ferromagnetic Fe nanowires on the [110] surface of superconducting Pb may support Majorana modes and topological superconductivity.
The theoretical prediction that proximity induced superconductivity in ferromagnetic (or half-metallic -- a half-metal is a perfect ferromagnet with complete spin-polarization at the Fermi level) wires might lead to a topological superconducting phase with the associated localized Majorana zero energy modes was made some years ago by several groups \cite{Lee_09, Duckheim, Chung_11,Takei}.
There was even an experimental report of the observation of long-range superconducting proximity effect through a ferromagnetic nanowire \cite{Samarth}.
Very recently, two theoretical papers \cite{Brydon,HBSTS} have predicted the possible existence of topological superconductivity in ferromagnetic nanowires lying on superconducting substrates using Shiba chain \cite{Brydon} and a one-dimensional nanowire\cite{HBSTS} model, respectively, closely mimicking the experimental system (i.e. Fe atoms on superconducting Pb) studied in Ref.~\citenum{Yazdani_14}.
Extensive recent theoretical work by different groups has suggested several different mechanisms which could lead to MBS-carrying TS in magnetic nanowires placed on superconducting substrates.
The earliest such mechanisms \cite{Nadj-Perge,Pientka,Glazman} modeled the nanowire as a chain of magnetic impurities in a spin-spiral phase.
The spin-spiral, following previous work \cite{Choy}, is used to mimic an effective spin-orbit coupling that would in turn lead to an effective triplet pairing superconducting proximity effect from the singlet superconducting substrate, exactly as in the existing semiconductor nanowire models of topological superconductivity \cite{Long-PRB,Roman,Oreg}.
The magnetic impurities were suggested to generate an array of Yu-Shiba-Rusinov \cite{YuShibaRusinov} bound states in the superconductor.
The combined effect of the superconductivity and spin-texture leads to an effective Kitaev chain model \cite{Kitaev-1D} that can support Majorana bound states under appropriate conditions \cite{Nadj-Perge,Pientka,Glazman}.
However, the theoretical plausibility of creating such a spin-spiral phase\cite{Loss, Franz} was debated, and it was shown that such spin spirals are unstable toward the formation of purely ferromagnetic or antiferromagnetic phases \cite{Lutchyn}.

The absence of a spin-spiral in the experimental system has led to the conjecture of an alternative mechanism involving the strong spin-orbit coupling of the Pb superconducting substrate itself contributing to topological superconductivity in the magnetic nanowire \cite{Lutchyn}.
The basic model, which has been studied in this context by several authors \cite{Yazdani_14, Brydon,HBSTS}, proposes that only the spin-triplet component of Cooper pairing, \textit{if any}, may be proximity-induced in a ferromagnetic wire from a spin-orbit coupled superconductor.
This mechanism has previously been proposed as an approach to topological superconductivity \cite{Duckheim} and also been invoked \cite{Takei}  to explain the long-range proximity-effect observed through ferromagnetic
nanowires \cite{Samarth}.
The mechanism of triplet proximity effect on a ferromagnetic wire arising from a spin-orbit coupled superconducting substrate has been studied in detail by three of us recently and shown to potentially support MBS-carrying topological superconductivity in the BDI chiral symmetry class \cite{HBSTS}.
Such a symmetry would suggest unsplit Majorana modes whenever the effective chemical potential in the ferromagnetic wire is positive.

In this context it is useful, particularly for later discussion of the experimental results \cite{Yazdani_14}, to briefly review two realistic theoretical models for topological superconductivity, and Majorana bound states, in one-dimensional (or quasi-one-dimensional) ferromagnetic nanowires (Fe in Ref.~\citenum{Yazdani_14}) lying on a two-dimensional surface of an underlying bulk superconductor (Pb in Ref.~\citenum{Yazdani_14}).  
One model, which we refer to as the Shiba model (or Shiba chain model), discussed in Refs.~\citenum{Pientka} and \citenum{Brydon} respectively for the helical and the ferromagnetic magnetic order in the wire, describes the magnetic atoms (Fe in Ref.~\citenum{Yazdani_14}) as essentially independent quenched classical magnetic impurities with little direct inter-atomic hopping along the chain, i.e. the one-dimensional band width of the chain is basically zero (or equivalently, vanishing inter-atomic hopping amplitude  $t$).  
The other model, which we refer to as the nanowire (or simply, the wire) model, introduced in Ref.~\citenum{HBSTS} for the ferromagnetic order in the magnetic chain, describes the ferromagnetic chain as strongly directly tunnel-coupled along the chain with considerable inter-atom hopping leading to one-dimensional bands of fairly large band-widths (or equivalently, large inter-atomic hopping amplitude $t$).  
These two models have been recently introduced and studied in the context of a ferromagnetic chain on a superconductor (i.e. the experimental system of Ref.~\citenum{Yazdani_14}) in Refs.~\citenum{Brydon} and \citenum{HBSTS} respectively.
The accompanying band structure calculations for the Fe chain on Pb presented in the experimental work \cite{Yazdani_14} indicate that the hoping term on the chain $t$ is of the order of $eV$ whereas the superconducting gap in Pb is of course of the order of $meV$.  
Given the large ratio between the hopping and superconducting parameters, as quoted in Ref.~\citenum{Yazdani_14}, the system can be modeled within the nanowire model.  
As we discuss later in this work, this fundamental inconsistency between different aspects of the results presented in Ref.~\citenum{Yazdani_14} remains unresolved with the reported induced topological gap being $10^{-4} eV$ and the observed strong lattice-level, $< 5 \; nm$, localization of the Majorana mode requiring an estimated superconducting gap of $\sim 1eV$.  
This basic incompatible dichotomy must be resolved (see below for a discussion of a recent work [60] in this context) before the observations of Ref.~\citenum{Yazdani_14} can be considered to be evidence for the existence of localized MBS in the Fe/Pb hybrid system.

A possible resolution of this apparent paradox, i.e. that of a very small proximity induced superconducting gap in the nanowire and a very small associated coherence length giving rise to strongly localized Majorana edge modes, has very recently been proposed in Ref.~\citenum{Peng} by Peng et al. using a zero-temperature theory of a helical magnetic chain on a superconducting substrate.  
The basic idea, originally studied in Refs.~\citenum{Sau_Robustness,Stanescu_Proximity}, for  2D heterostructures made of semiconductor/superconductor \cite{Sau_Robustness} or topological-insulator/superconductor \cite{Stanescu_Proximity} sandwich systems, is that the presence of the superconducting substrate could substantially renormalize the parameters of the effective induced superconductivity, leading to a situation where the effective parameters determining the proximity superconductivity are radically different from those applying to an isolated p-wave superconducting system (i.e. without any substrate).  
In particular, Peng et al. explicitly show\cite{Peng}, by using a minimal mean-field-theoretical model of a linear helical chain of Anderson magnetic adatom impurities on a bulk superconductor, that a strong velocity renormalization in the wire induced by the substrate could lead to a small Majorana localization length in spite of the wire having a small induced superconducting gap.  
If this physics is indeed operational in the experimental system of Ref.~\citenum{Yazdani_14}, then the lowering of temperature should show considerable sharpening of the zero-bias peak in future experiments even if the induced gap itself remains small.  Our interest in the current work is, however, a theoretical modeling of the high-temperature situation (with the induced gap being comparable to the temperature) as studied in Ref.~\citenum{Yazdani_14}, and the precise Majorana localization itself is not particularly important for considering the high-temperature situation as we show in the finite-temperature results presented in this paper.  
One issue directly resolved by Peng et al. \cite{Peng} is that the ferromagnetic nanowire model considered in this work and the Shiba model are {\em not} separated by any quantum phase transition -- they are two extreme aspects of the same underlying crossover physics akin to the BCS-BEC crossover in superfluidity/superconductivity with the appropriateness of each description being determined by underlying experimental parameters.  
We mention, however, that the key parameter underlying the strong Majorana localization established in Ref.~\citenum{Peng} is the strength of the tunneling between the substrate and the adatom chain, and experimentally there is no information on how strong this tunneling is. 
When this tunneling is weak (strong), the ferromagnetic wire (Shiba chain) model applies.  In the current work, we primarily follow the ferromagnetic nanowire model since this is the description used in Ref.~\citenum{Yazdani_14} itself to describe the data, although we specifically show (see Figs.~18 and 19) that at the high experimental temperatures used in Ref.~\citenum{Yazdani_14}, the tunneling conductance measured experimentally would look very similar in either the ferromagnetic or the Shiba model, and the physics of strong Majorana localization would become relevant only if the experimental temperature is lowered to well below (less than half at least) the induced superconducting gap in the nanowire.

We emphasize again that the Shiba and nanowire models are two extreme and complementary aspects of the same underlying physics (i.e. the physics of a linear array of ferromagnetic magnetic atoms lying on a 2D or 3D superconducting substrate with strong spin-orbit coupling in the environment -- a 1D magnetic chain on a 2D or 3D superconducting substrate in the presence of strong spin-orbit coupling), and there is no phase transition between the two situations.
Indeed the models can be understood as two limiting cases where the weak (strong) hopping between d-band orbitals corresponds to the Shiba (nanowire) regime\cite{Peng}. 
With our main interest in the current work being the role of temperature (compared with the induced superconducting gap in the wire), it should not matter which model is used, although the nature of the Majorana mode localization scales are very different in the two models\cite{Peng} depending on the crossover parameter given by the tunneling hybridization scale between the superconductor and the magnetic chain.
An extremely important physical parameter in the experiment of Ref.~\citenum{Yazdani_14} is the very high experimental temperature ($T\sim 1.2K$) which is comparable to the estimated induced topological superconducting gap ($\Delta \sim .12 \;meV$) extracted in Ref.~\citenum{Yazdani_14}. 
The fact that $k_bT \sim \Delta$ in the experiment makes any discussion of a precise zero-energy Majorana mode quite meaningless because at such high temperatures, the Majorana response will be broadened over the whole subgap regime (or perhaps even above the gap) with the Majorana signal indistinguishable from any ordinary fermionic subgap state. 
We also point out that the experiment of Ref.~\citenum{Yazdani_14} does not actually observe any obvious superconducting gap in the ferromagnetic nanowire, and the evidence for the existence of any topological gap is indirect.  
In fact, the instrumental energy resolution in Ref.~\citenum{Yazdani_14} is also of the order of the topological gap and the temperature, making any discussion of possible weak subgap features as representing anything definitive somewhat premature.
Given that the experiments in Ref.~\citenum{Yazdani_14} are carried out at temperatures equal to the induced superconducting gap using an experimental instrumental resolution also comparable to the gap, we believe that any subgap features would manifest very broad and very weak differential tunneling conductance peak consistent with the observations, and only a lowering of the temperature in future experiments can distinguish possible zero-bias Majorana features from fermionic subgap features.

Our goal in the current paper is a detailed numerical modeling of temperature effects in the ferromagnetic nanowire-superconductor hybrid system in the presence of strong spin-orbit coupling in order to provide qualitative (and semiquantitative) insight into the physics of the system studied in Ref.~\citenum{Yazdani_14}.
The current work is an extension of Ref.~\citenum{HBSTS} carried out in the context of the putative experimental MBS observation claimed in Ref.~\citenum{Yazdani_14} in order to provide a detailed critical comparison between theory and experiment, which is necessary since the rather strong claim of a direct observation of Majorana modes must be thoroughly validated from all possible perspectives.  It may be useful in this context to emphasize that the recent spurt in the experimental MBS activity, including both the earlier work on semiconductor (InSb and InAs) nanowires \cite{Mourik,Deng,Weizman,Rokhinson,Churchill,Finck} and the very recent work on Fe nanowires\cite{Yazdani_14}, is completely dependent on theoretical predictions and analyses for its validation since the observations themselves involving tiny zero-bias tunneling conductance peaks at low temperatures in rather complex hybrid systems are remarkably unremarkable, becoming noteworthy only because theories specifically predicted that such zero-bias tunneling peaks should exist in these specific hybrid structures as MBS signatures.
In particular, Ref.~\citenum{Long-PRB} not only predicted the existence of the Majorana bound states in semiconductor-superconductor hybrid structures, specifically laying out the type of structures (and the materials) experiments should use, but also carried out realistic calculations showing that the resulting MBS-induced zero-bias tunneling peaks should have a small height (because of finite temperatures, tunnel barrier heights, and wire lengths) compared to the expected quantized value \cite{Sengupta-PRB-2001} associated with the perfect Andreev reflection anticipated for MBS.
This early paper\cite{Long-PRB} also specifically suggested the use of STM in order to look for topological zero energy Majorana excitations in hybrid systems as has eventually been accomplished in Ref.~\citenum{Yazdani_14} following the later suggestion in Ref.~\citenum{Nadj-Perge} of using an STM coupled specifically with a magnetic chain on a superconductor.

While topological superconductivity in the chiral symmetry class has been established for ferromagnetic wires with a single spatial orbital per atom, the number of Majorana modes arising from such a model is limited to two.
On the other hand, the band-structure calculation for the experimentally realistic system \cite{Yazdani_14} suggests that the number of channels in the wire can be significantly enhanced by the presence of multiple orbitals per atom and multiple atoms along the diameter of the chain.
In this work, we consider a multichannel generalization of the FM heterostructure and its topological properties starting from the nanowire model of Ref.~\citenum{HBSTS}.
Non-trivial zero-bias phenomena appear across a broad range of parameters in contrast to the fine tuning necessary for a non-trivial topological phase in class D \cite{Sau-Generic,Tewari-Annals,Long-PRB,Roman,Oreg} or DIII systems \cite{Sato_11,Law_12,Deng_12,Keselman_13,Zhang_13,Deng_13,Law_14,Flensberg_13,Nakosai_13,Dumitrescu_TR}.
Within this framework, manipulating the system width (i.e. coupling parallel magnetic chains) enhances or reduces the zero-bias conductance peak (ZBCP) height accordingly.
Manipulation of the zero-bias conductivity tuned by the width of the magnetic chain would be a direct signature of the chiral class BDI topological superconductors.
Additionally, we calculate spatially resolved scanning tunneling conductance profiles including effects of finite temperature and finite size of the wire (as well as the finite tunnel barrier effects) which are experimentally accessible by STM.
Finally, we show that the fractional Josephson effect maintains its $4 \pi$ periodicity in phases supporting multiple spatially overlapping MBS and comment on how the Josephson current may be enhanced in the presence of Majorana multiplets \cite{Kitaev-1D, Roman} for a definitive conclusion regarding the existence of MBS in the ferromagnetic nanowire system of Ref.~\citenum{Yazdani_14}. 

It may be useful to point out the connection between MBS in much-studied semiconductor nanowire systems with that in the new platform of interest in Ref.~\citenum{Yazdani_14} involving ferromagnetic nanowires.  
Although it may appear at first sight that the two systems are completely distinct, from a theoretical perspective the MBS in the ferromagnetic wires are described by essentially the same theory as developed earlier for the semiconductor nanowires in Refs.\citenum{Long-PRB,Roman,Oreg}, provided that one is in the nanowire limit of large inter-atomic hopping along the chain (and not in the Shiba limit), and that one is in the limit of the spin splitting (induced in the semiconductor case by an external magnetic field or by a proximate exchange splitting) being very large (much larger than the other energy scales in the problem including the spin-orbit coupling energy, the Fermi energy, and the superconducting gap in the ferromagnetic wire case).
In this large spin-splitting limit, the semiconductor system is also essentially an effective ``half-metallic ferromagnet" exactly as the Fe wire studied in Ref.~\citenum{Yazdani_14} is claimed to be.
In the semiconductor nanowire case also, the topological superconducting phase will be generic in this very large spin-splitting limit since the chemical potential would by definition be in the single spin polarized subband, with the superconducting gap being smaller than the spin splitting.
Thus, the distinction made between semiconductor nanowires and magnetic nanowires with respect to topological superconductivity is a distinction without much difference, since one can take the existing theory for the semiconductor nanowire and obtain all the necessary formula for the ferromagnetic wire case by assuming the spin-splitting to be by far the largest energy scale. (We provide the details on this connection to the semiconductor nanowire system in Appendix A).
We emphasize, however, the obvious fact that although the ferromagnetic wire case can be thought to be a limiting situation (i.e. very large spin-splitting limit) of the semiconducting nanowire Majorana theory, the two experimental systems (namely, the semiconductor nanowire in the presence of an external magnetic field inducing spin-splitting and the ferromagnetic nanowire with its spontaneous exchange-driven spin-splitting) are, of course, completely different physical platforms from an experimental perspective utilizing totally different materials and measurement techniques.
The emergence of this ferromagnetic nanowire platform \cite{Yazdani_14}, in addition to the already existing semiconductor nanowire platforms \cite{Mourik,Deng,Weizman,Rokhinson,Churchill,Finck}, is therefore an exciting new development in the search for MBS and topological quantum computation.

\section{Experimental Setup and Theoretical Model}

\begin{figure}[tb!]
\centering
\includegraphics[width=\columnwidth]{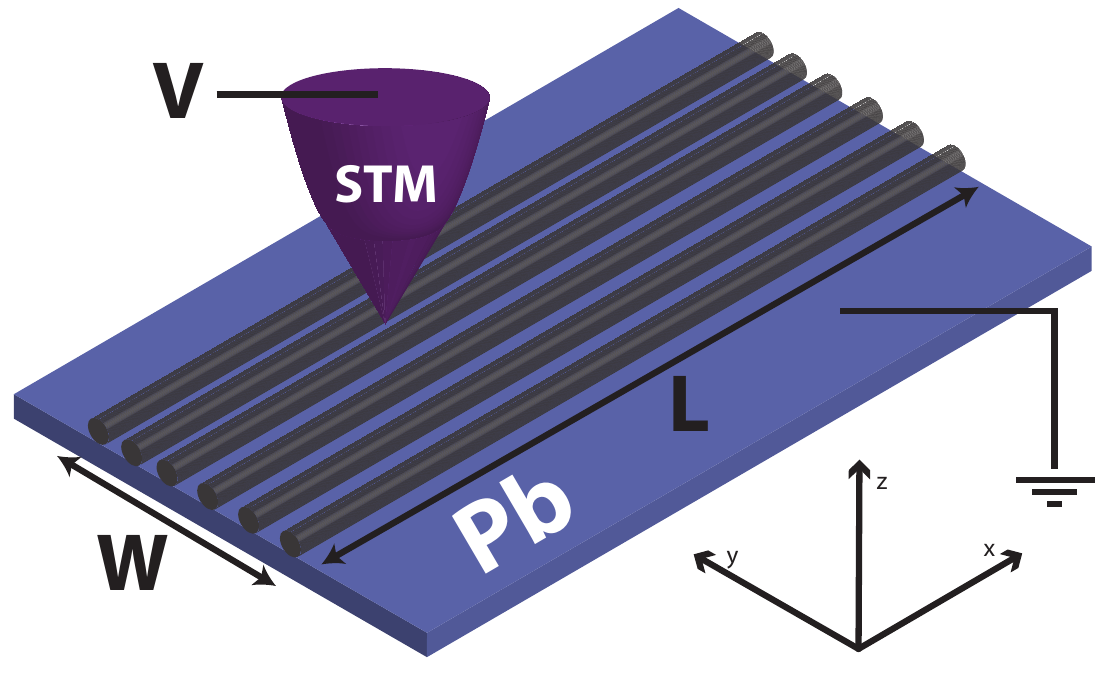}
\caption{(Color online) a) Schematic diagram of the proposed heterostructure involving a series of ferromagnetic quantum wires (gray), with large intrinsic magnetization M, deposited on top of a spin-orbit coupled s-wave superconductor such as Pb (blue substrate).
Spin singlet and triplet pairing potentials are proximity induced in the FM wires due to the strong spin orbit coupling and inter-orbital mixing in the superconductor.
A STM probe, at coordinate x, measures the spatial dependence of the differential conductance along the longitudinal axis.}
\label{fig:Schematic}
\end{figure}

Taken very close to a sample surface, an STM can be used as an electrode to measure transport properties (Fig.~\ref{fig:Schematic}).
A movable scanning point contact tunneling experiment is potentially very useful in investigating the edge character of Majorana zero modes since the STM is particularly well-suited in measuring the local density of states.
This idea, originally proposed in Ref.\citenum{Long-PRB}, is rather impressively implemented in the highly demanding spin-polarized STM measurements presented in Ref.~\citenum{Yazdani_14}.
Provided the electrical contact is good between the ferromagnet (Fe nanowire) and the superconducting substrate (Pb), Cooper pairs will leak into the ferromagnet, thereby proximity inducing superconductivity in the nanowire.
We model a finite quantum wire with dimensions $L_y \ll L_x \equiv L$  by considering a $N_x \times N_y$ site square lattice with unit spacing.
The effective Hamiltonian for the topological superconductor, within the pure nanowire limit, is $H_{TS} = H_{t} + H^{S=0}_{\Delta} +H^{S=1}_{\Delta} +H_{Z}$ where
\begin{equation}
\begin{aligned}
H_{t} 	& = \sum_{\<ij\>\sigma} t \left[ c^{\dagger}_{i \sigma} c^{ }_{j \sigma} + \text{H.c.} \right] - \sum_{j \sigma} \mu_j c^\dagger_{j \sigma} c_{j \sigma} \\
H^{S=0}_{\Delta}& = \sum_j \Delta_s c^\dagger_{j \uparrow}c^\dagger_{j \downarrow}+\text{H.c}\\
H^{S=1}_{\Delta}& = \sum_{j \sigma} i \Delta_p (c^\dagger_{j \ua}c^\dagger_{j+1 \ua}-c^\dagger_{j \da}c^\dagger_{j+1 \da})+\text{H.c}\\
H_{Z} 	& = \sum_{j \sigma \sigma '} c^{\dagger}_{j \sigma}( \bm{V}\cdot \bm{\sigma} )_{\sigma \sigma'} c^{ }_{j \sigma'}
\end{aligned}
\label{eq:TBModel}
\end{equation}
Here $c^\dagger_j$ is the electronic creation operator for site $j$, $\<ij\>$ indicates nearest neighbor sites, $\bm{\sigma}=(\sigma_x,\sigma_y,\sigma_z)$ is the vector of Pauli matrices, $t$ is the hopping amplitude in the nanowire, and $\mu$ is the chemical potential.
The superconducting pairing is a mixture of singlet and triplet terms.
For the triplet pair potential term we have taken a Cooper pairing with spin projection $S_x=0$, which is the familiar equal-spin-pairing $\Delta_{\da \da}=-\Delta_{\ua \ua}$.
The Zeeman spin-splitting due to an internal magnetization in the ferromagnet is $\bm{M}$ is $\bm{V}=g \mu_B \bm{M}=(V_x,V_y,V_z)$ where $g$ and $\mu_B$ are the Lande g-factor and Bohr magneton respectively.
We note that our effective Hamiltonian, as given in Eq.~\ref{eq:TBModel}, describes the TS phase of the ferromagnetic nanowire assuming that the degrees of freedom of the underlying superconducting substrate (Pb in Ref.~\citenum{Yazdani_14}) have been integrated away with Eq.~\ref{eq:TBModel} now describing only the electrons in the Fe magnetic wire.
We refer to Ref.~\citenum{HBSTS} for the details on how to obtain Eq.~\ref{eq:TBModel} which is our starting point in the current work.
We note that in this context our effective model, derived from Ref.~\citenum{HBSTS} which should be consulted for the details, describes only the ferromagnetic nanowire, hiding all information about the underlying superconducting substrate with the parameters for the spin-orbit coupling, the bulk superconducting gap of the substrate, the hopping amplitude of Cooper pairs between the substrate and the nanowire inducing the singlet and triplet proximity effect, etc. being implicitly contained in the induced superconducting pair potentials $\Delta_s$ and $\Delta_p$, which we use as phenomenological parameters to be obtained from the experimental measurements themselves. 
Our goal here is to obtain the phenomenological consequences of the minimal topological nanowire model (i.e. Eq.~\ref{eq:TBModel}) for the ferromagnet/superconductor hybrid system to make observable predictions and to carry out comparison with the existing data.
We also ignore all nonessential complications such as the number of orbitals per Fe atom and the effective width of the wire, and so on which can be absorbed in the multichannel generalization we consider below (i.e. the W-parameter denoting the number of active wire channels as described below).  Our goal here is to utilize the minimal model and work out its implications in great details.
Our Eq.~\ref{eq:TBModel} serves as the minimal model for the experimental system of Ref.~\citenum{Yazdani_14} in the current work.

Throughout this work we fix all of our parameters relative to the hopping integral $t$ in the nanowire.
To begin and to establish our general results, we use $\Delta_s=\Delta_p=t/10$, $\bm{V}=V_z =2.0 t, L=100$ while $W$ (the number of transverse channels) and the chemical potential $\mu$ are allowed to vary.
For simplicity and numerical convenience, we will choose $\Delta_s=\Delta_p$. 
Choosing two such values $\Delta_s=\Delta_p=0.1 t$ and $\Delta_s=\Delta_p=0.01 t$ (both of which are orders of magnitude larger than reported in Ref.~\citenum{Yazdani_14} assuming $t \sim 2 \; eV$ as given in Ref.~\citenum{Yazdani_14}) will allow us to estimate the order of magnitude of parameters such as the Majorana decay length (see Fig.~\ref{fig:psi}).
We will present our numerical results for a few values of $L$,$\Delta$, and $T$ including $T=0$ results (for the sake of comparison). 
A discussion concerning experimentally realistic parameters and their effect on the measured tunneling conductance is left to a later section.
We note that this choice of generic parameters incorporates the half-metallic character of the ferromagnetic wire since only one spin subband is occupied for a large range of chemical potential values keeping the system in the topological phase without any additional fine-tuning of parameters.
Solving Eq.~\ref{eq:TBModel} directly numerically we find zero energy Majorana states which are localized at each end of the wire.
The evolution of the low energy spectrum as a function of the chemical potential, as well as a function of the number of zero energy modes, is presented in the bottom panel of Fig.~\ref{fig:Bands}.
To understand how Eq.~\ref{eq:TBModel} realizes an integer number of Majorana zero modes we analyze the topological properties of this model in the next section.

\section{Topological Properties and Quantum Phase Transitions}
\label{sec:topo}

\begin{figure}[t!]
\centering
\includegraphics[width=\columnwidth]{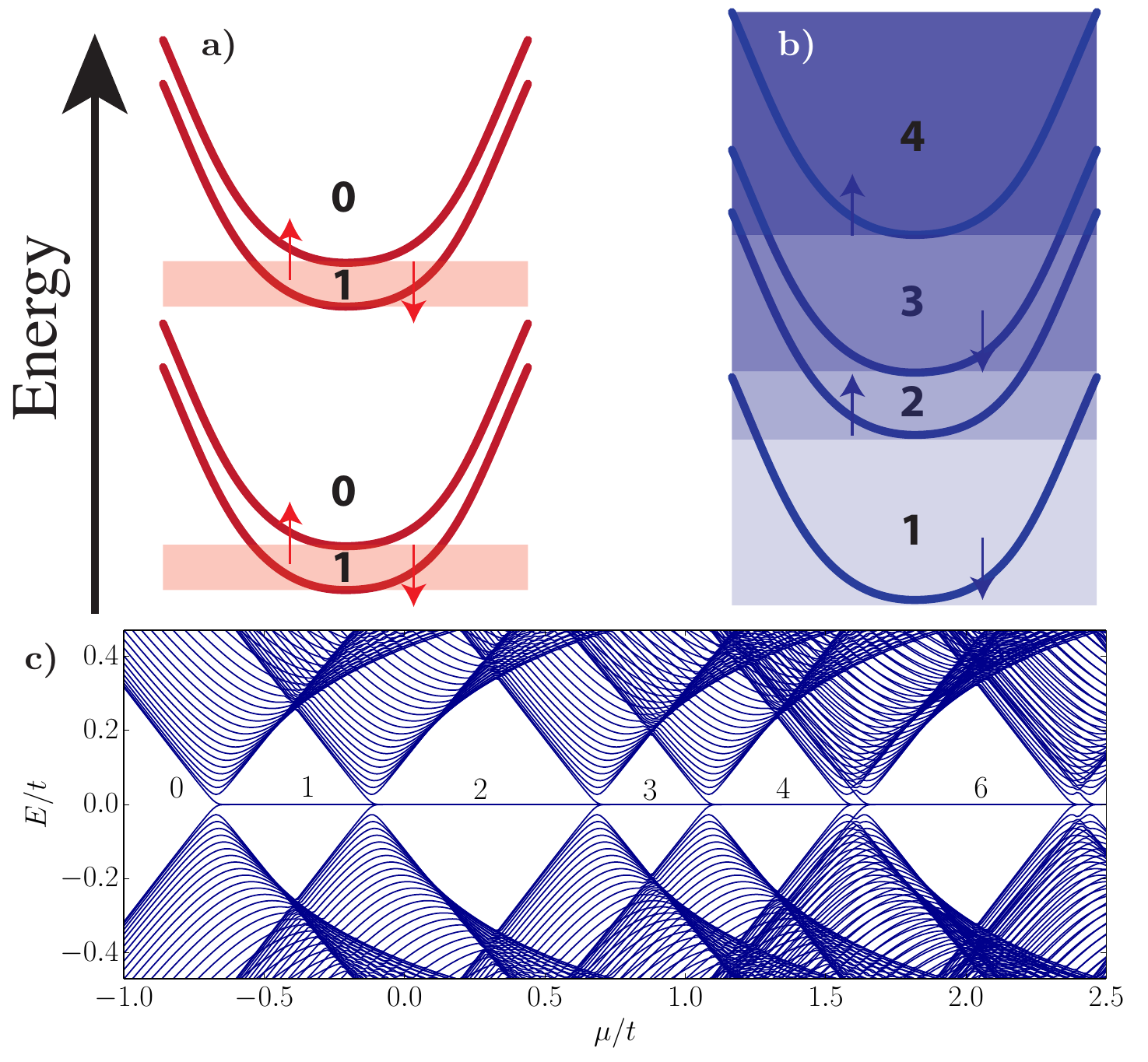}
\caption{(Color online) Comparison of the topological properties of the class D spin-orbit coupled semiconducting nanowire (red) and our ferromagnetic system (blue).
In panel (a) an externally applied magnetic field induces a Zeeman splitting between the originally degenerate spin bands.
This system is topological non-trivial if an odd number of bands is occupied and Cooper pairs are supplied by a nearby superconductor.
When the Fermi energy lies in the shaded region, the class D $\mathbb{Z}_2$ invariant non-trivial and a single Majorana bound state emerges at each end.
b) Normal state band structure for a multichannel ferromagnetic wire (see Fig.~\ref{fig:Schematic}).
In the presence of proximity induced p-wave pairing, the FM is promoted to the topological class BDI, which is characterized by a non-trivial $\mathbb{Z}$ invariant for a generic band occupancy.
Additionally, the large intrinsic magnetization provides a broad non-trivial parameter regime in which a non-trivial topological state persists even if the chiral symmetry is broken, say, by a second Zeeman field perpendicular to the magnetization (BDI $\rightarrow$ D).
In this case, the shaded regions with an odd $\mathbb{Z}$ invariant remain non-trivial while the others become trivial.
(c) Low energy quasiparticle spectrum as a function of the chemical potential in the ferromagnetic wire.
As $\mu$ increases, the number of Majorana zero modes at each end of the FM wire increases by one following each gap closing.}
\label{fig:Bands}
\end{figure}

According to the Altland-Zirnbauer classification scheme\cite{AZ}, free fermion systems are characterized by their dimensionality as well as by the presence and the sign of anti-unitary symmetries.
There are ten topological classes in total and five of them are non-trivial (i.e. a non-trivial topological invariant can be defined) for a given dimension.
The two anti-unitary symmetries used are time-reversal symmetry (TRS) and particle-hole symmetry (PHS), with the latter often being referred to as the charge conjugation symmetry.
Denoting the TRS and the PHS operators by $\Theta$ and $\Xi$ respectively, the anti-unitary symmetries are present when the following reality conditions are satisfied: $\Theta H \Theta^{-1}  =  U_\Theta H ^{*} U_\Theta^{\dagger}  = +H$ and $\Xi H \Xi^{-1}  =  U_\Xi H^{*} U_\Xi^{\dagger}  = - H $.
Here $U_{\Theta,\Xi}$ denote the unitary part of the TR and PH operators.
A system is chiral invariant (or sublattice symmetric) when both TR and PH are present and is given by the unitary operator ${\cal{S}}=\Theta \cdot \Xi$.
The classification triplet $(T,C,S)=(\Theta^2,\Xi^2,{\cal{S}}^2)$ is used to index each symmetry class, where TR and PH operators can square to $\pm1$ and the chiral operator is restricted to ${\cal{S}}^2=+1$.
We write ${\cal{O}}^2=0$ if an operator is not present.

Invariants are generally formulated in terms of the bulk Hamiltonian's topology, so we now look at a strictly 1D version of $H_{TS}$.
Fourier transforming Eq.~\ref{eq:TBModel} with a single spatial channel (i.e. no transverse hopping), the momentum space Bogoliubov-de Gennes (BdG) Hamiltonian becomes $H=\sum_{k} \Psi_k^{\dagger} H (k) \Psi_k^{ }$ where
\begin{eqnarray}
H (k)  & = & (-2t \cos(k) -\mu) \sigma_0 \tau_z \\ \nonumber
          & + & \left[ \Delta_s \sigma_0 + \Delta_p \sin(k) \bm{d} \cdot \bm{\sigma} \right]      \tau_x \\ \nonumber
	  & + & \bm{V} \cdot \bm{\sigma} \tau_0.
\label{eq:HBDG}
\end{eqnarray}
Here $k \equiv k_x$ is the one-dimensional crystal momentum and $\Psi_{k}=(c_{k\uparrow},c_{k\downarrow},c_{-k\downarrow}^{\dagger},-c_{-k\uparrow}^{\dagger})^{T}$ is our four component Nambu spinor which acts on the particle-hole $(\bm{\tau})$ and spin spaces $(\bm{\sigma})$.
In our calculations we use $\bm{d}=(1,0,0), \bm{V}=(0,1,0)$ but leave $\bm{d},\bm{V}$ in Eq.~\ref{eq:HBDG} to highlight the generic properties of the various symmetry classes.
Given our choice of basis, the anti-unitary TR and PH operators have the matrix structure $\Theta = i \sigma_y \tau_0 {\cal{K}}$ and $\Xi = \sigma_y \tau_y {\cal{K}}$ where ${\cal{K}}$ is the complex conjugation operator.

In momentum space, the reality conditions for Bloch Hamiltonians are \cite{Hasan-Kane}
\begin{eqnarray}
\label{eq:momentum_reality}
\Theta H (k) \Theta^{-1} & = & + H (-k)  \\ \nonumber
\Xi H (k) \Xi^{-1} & = & - H (-k).
\end{eqnarray}
PH symmetry emerges from the BCS mean-field theory and is intrinsic to all BdG Hamiltonians, so in the absence of any additional symmetries $(T,C,S)=(0,1,0)$.
This triplet corresponds to the topological class D, which is characterized by a $\mathbb{Z}_2$ topological invariant in $d=1$.
Recall, $\mathbb{Z}_2 \equiv \mathbb{Z}/2\mathbb{Z}$ is the cyclic quotient group with two elements $\left\{ 0,1 \right\}$.
The topological invariant for class D systems is given by Kitaev's Majorana number, which is defined as \cite{Kitaev-1D} ${\cal{M}} = \text{sgn}\left[ Pf(\widetilde{A}(0)) Pf(\widetilde{A}(\pi)) \right]$ where $\widetilde{A}$ is the momentum space Hamiltonian written in a skew symmetric form, determines when the system is topologically non-trivial.
The MBS-carrying spin-orbit coupled semiconductor-superconductor heterostructure proposal\cite{Sau-Generic,Tewari-Annals,Long-PRB,Roman,Oreg} belongs to the topological class D for the most general types of spin-orbit coupling. 
However, the specific models studied in the original proposals\cite{Sau-Generic,Tewari-Annals,Long-PRB,Roman,Oreg} assumed that the spin-orbit direction was perpendicular to the Zeeman coupling. 
These models are therefore in a more restricted BDI class that will be discussed further at the end of this section because of its relevance to the ferromagnetic wire model.
In this system, a Zeeman field splits degenerate spin-orbit coupled bands. 
The goal of this splitting is to remove a single Fermi surface thus rendering the system effectively spinless.
Typically, the Zeeman splitting is small compared to the bandwidth, resulting in a small non-trivial topological parameter range (see Fig.~\ref{fig:Bands}).
In this case, the difficult task of fine tuning the chemical potential, by using gate electrodes for example \cite{Mourik,Deng,Weizman,Rokhinson,Churchill,Finck}, may be necessary to achieve a non-trivial topological state if the chemical potential lies near half filling for a subband.
As emphasized at the end of the Introduction, however, we are free to take the very large spin-splitting (i.e. very large $V_z$) limit of the semiconductor model (although this would not be a particularly physically relevant model for semiconductors per se, it is a perfectly allowed theoretical limit), which then coincides with the current ferromagnetic wire situation of interest to the experimental system in Ref.~\citenum{Yazdani_14} (see Appendix A for the details).

If Zeeman splitting is absent in Eq.~\ref{eq:HBDG}, then the first reality condition from Eq.~\ref{eq:momentum_reality} is satisfied.
Using $\Theta = i \sigma_y \tau_0 {\cal{K}}$ we see that  $T=-1$ so a class DIII TR invariant system is characterized by the triplet $(T,C,S)=(-1,1,1)$.
Class DIII is characterized by a $\mathbb{Z}_2$ topological index which is related to a Kramers polarization \cite{Budich-Ardonne2} similar to the way the class D invariant is related to the electric polarization of the wire \cite{Hasan-Kane}.
If $H$ belongs to this class, any hybridization between time-reversed Majorana zero modes is forbidden by Kramers degeneracy, and each end of the wire hosts a perfectly degenerate Majorana Kramers pair.

In addition to the two classes discussed above, superconducting systems can belong to the topological class BDI.
Note that there exists the chiral operator ${\cal{S}}=\hat{d} \cdot \bm{\sigma} \tau_y$ which anti-commutes with the Hamiltonian in Eq.~\ref{eq:HBDG}; note in the TR invariant case the chiral operator is ${\cal{S}}_{DIII}=\sigma_0 \tau_y$.
This unitary chiral operator must be a product of two anti-unitary operators, one of which is $\Xi$.
By simple algebra, one can show that our missing operator is ${\cal O}=(\hat{d}\cdot\hat{y}+i(\hat{d}\times\hat{y})\cdot\bm{\sigma}){\cal K}$ so that ${\cal O}^2=1$ (i.e. $T=+1$) and that $H(k)$ satisfies ${\cal O} H (k) {\cal O}^{-1} = + H (-k)$.
We continue to call this operator ${\cal O}$, even though it leads to the same reality condition as $\Theta$, in order to distinguish it from the usual time reversal symmetry.
A crucial difference between classes BDI and D/DIII is that the former is characterized by an integer $\mathbb{Z}$ invariant.
Because the invariant can take any integer value, multiple spatially overlapping MBS can coexist in contrast to class D systems where localized zero-energy anyonic MBS hybridize into conventional finite-energy fermionic quasiparticle states.
As illustrated in Fig.~\ref{fig:Bands} panel (b), a BDI chiral system is non-trivial for a generic parameter range.
We numerically diagonalize and plot the low energy quasiparticle spectrum for $H_{TS}$ as a function of the chemical potential in Fig\ref{fig:Bands} panel (c).
The Majorana occupancy grows when $\mu$ increases and successive higher energy bands are occupied. Therefore for any generic chemical potential one expects a non-trivial topological state with end-localized zero energy MBS.

Composing the two reality conditions in Eq.~\ref{eq:momentum_reality} we see that the chiral operator satisfies $\left \{ {\cal{S}}, H (k) \right \} = 0$.
This anti-commutation relation implies that in the eigenbasis of ${\cal{S}}$ the Hamiltonian is off-diagonal,
\begin{equation}
\label{eq:chiral_form}
H' (k) =
\left(
\begin{array}{cc}
0 & A(k)  \\
A^{\dagger}(k)  & 0
\end{array}
\right).
\end{equation}
Here we have used $U$ to represent the unitary transformation matrix between the original and the chiral basis.
For a single channel $A(k)$ is a $2\times2$ complex Hermitian matrix whose determinant $D(k) \equiv Det(A(k))$ is generally complex.
Obviously the complex phase $\exp[i \theta(k)]=D(k)/|D(k)|$ lies on the unit circle and we have established a mapping from the Brillouin zone ($S^1$ in 1D) to $U(1)$ .
The fundamental group $\pi_1 (U(1))=\mathbb{Z}$ is well defined here so that we may write the topological winding invariant as \cite{TS_BDI},
\begin{equation}
\label{eq:winding}
{\cal{W}}=\frac{1}{2\pi} \int_{0}^{2\pi} \arg D(k) dk.
\end{equation}
The integer ${\cal{W}}$ counts the number of times the complex argument $\theta(k)$ winds about the origin in the complex plane and is invariant under smooth deformations.
In other words, ${\cal{W}}$ can change only if the winding curve $D(k)$ passes through the origin.
However, by looking at the from of Eq.~\ref{eq:chiral_form} we know that the $k$-point where $D(k)$ vanishes constitutes a gap closing with a concomitant topological quantum phase transition.
Note that the DIII chiral operator is odd under time-reversal symmetry, $\left\{{\cal{S}}_{DIII},\Theta\right\}=0$ which implies that the end modes with chiral charge $+1$ are compensated by an equal number of modes with charge $-1$.
Therefore, while this procedure may be mathematically well defined, it is trivial in the sense that the net DIII chiral topological charge always vanishes.

The winding number defined in Eq.~\ref{eq:winding} can also be used to calculate the chiral topological invariant for multichannel wires \cite{Qu}.
The quasi-one-dimensional Hamiltonian used in this procedure is one in which a Fourier transform has been performed along the longitudinal x-direction, but not along the y-direction.
Using $l ,l'\in \left[0,W\right]$ to indicate the y-coordinate, we write $H_{TS}=\sum_{k l l'} \Psi_{k l}^{\dagger} (H(k) \delta_{l,l'}+H^{\perp}_{l,l'}) \Psi_{k l'}$  where $H^{\perp}_{l,l'} = - t \sigma_0 \tau_z  (\delta_{l,l'+1}+\delta_{l,l'-1})$.
We use the procedure outlined above, where $A(k)$ is now a $2 W \times 2 W$ dimensional matrix and multiple Bloch bands can now be mapped to $U(1)$ by the determinant function.
The result is sketched in Fig.~\ref{fig:Bands} panel (b).
As the chemical potential increases and higher bands are filled, the gap closing occurs in the spectrum of $H_{TS}$ and the corresponding topological invariant ${\cal{W}}$ increases by unity.
We emphasize that when the chemical potential is in the lowest spin-split band (Fig.~\ref{fig:Bands}), the topological phase is generically present in this half-metallic FM situation since the spin-splitting is much larger than the induced superconducting gap.
We shall now discuss the experimental signatures which are a consequence of our model.

\section{Scanning Tunneling Differential Conductance}

Consider an STM brought close to the surface of the multichannel FM wire described by Eq.~\ref{eq:TBModel} (see Fig.~\ref{fig:Schematic}) .
The STM tip weakly couples to FM wire orbitals through a small hopping integral $ H_{STM} = \sum_{\sigma} t' (c^{\dagger}_{s \sigma} d^{ }_{s \sigma} + \text{H.c.} )$.
Here $d$ annihilates electrons at the STM tip which we take to be three sites wide and centered the x-coordinate $s = (x-1, x, x+1)$.
We will parametrize the tunneling barrier at the STM tip (which determines the size of the zero bias tunneling peak at finite temperatures \cite{Sengupta-PRB-2001,Long-PRB}) by the single parameter $t'$ for simplicity-- typically $t' \ll t$ in the STM set up of Ref.~\citenum{Yazdani_14}.
A potential difference $V$ is now applied between STM and drain (i.e. the grounded superconductor on which the FM has been deposited).
We will now set up a scattering matrix formalism to calculate the differential conductance through the FM wire, in order to experimentally detect the MBS.
Within this approach we model the STM, which is the first scattering lead, as a normal metal electron reservoir biased at a variable electrochemical potential $\mu_N+eV$ measured relative to the superconducting Fermi energy.
Our quasi-one-dimensional FM wire (Eq.~\ref{eq:TBModel}) acts as the scattering region and the second lead is the grounded electron drain which is held at chemical potential $\mu_N$.
We adopt a BTK perspective \cite{BTK} in assuming that equilibrium Fermi distribution functions determine the incoming quasiparticle occupancy levels.
Here, $\psi_{in}=(\psi^{S}_{in},\psi^{D}_{in})^T$ are plane waves originating deep within the semi-infinite lead STM and drain leads which are described by the Fermi functions $f(E-eV)$ and $f(E)$ respectively.
Note, in general $\psi^{S,(D)}$ is an $N (M)$ component spinor given $N (M)$ occupied channels in the STM (drain) lead.
For a quantum coherent process we can relate the outgoing modes to the incoming modes by the scattering matrix $\psi_{out} = \hat{S} \psi_{in}$ where
\begin{equation}
\hat{S} = \left(
\begin{array}{cc}
 r & t' \\
 t & r' \\
\end{array} \right).
\label{eq:SMatrix}
\end{equation}
Here $r$ is a $4N \times 4N$ matrix consisting of complex reflection coefficients between all the occupied incoming STM channels.
Likewise $r'$ is the reflection matrix for the drain and $t,t'$ are the transmission coefficient matrices connecting the two leads. 
(Note that we use the same notations $t, t'$ to denote the transmission matrix elements for the leads as what were used to define the tunneling amplitudes in defining the basic Hamiltonian, but there is no scope for any confusion here since the transmission matrix elements $t, t'$ only appear in Eq.~\ref{eq:SMatrix} above in defining the $\hat{S}$-matrix and in our numerical work and nowhere else in the text below.)

In the presence of a proximity induced superconducting gap, single electrons cannot tunnel from the STM to the FM for low bias-voltages $V \ll \Delta$.
As a result, all of the flowing current in the subgap regime is generated through the Andreev reflection process in which excess Cooper pairs are created and simultaneously the incident electrons are converted into holes.
The reflection matrix can be written as
\begin{equation}
r = \left(
\begin{array}{cc}
 r_{ee} &r_{eh} \\
 r_{he} & r_{hh} \\
\end{array} \right).
\end{equation}
where $r_{ee}$ ($r_{eh}$) refers to the normal (Andreev) reflection submatrix.
At low bias voltages the differential conductance, proportional to the transmission probability at a given energy E, is expressed in terms of the STM reflection matrices as \cite{San-Jose}
\begin{equation}
\label{eq:dIdV}
\frac{dI(V)}{dV}=\frac{e^2}{h} \left[ N - Tr(r_{ee}^{\dagger} r_{ee}) + Tr(r_{eh}^{\dagger} r_{eh}) \right]_{E=V}.
\end{equation}
We generate the scattering coefficients numerically using the Kwant \cite{Kwant} numerical package.

\subsection{Results}

\begin{figure}[t!]
\centering
\includegraphics[width=\columnwidth]{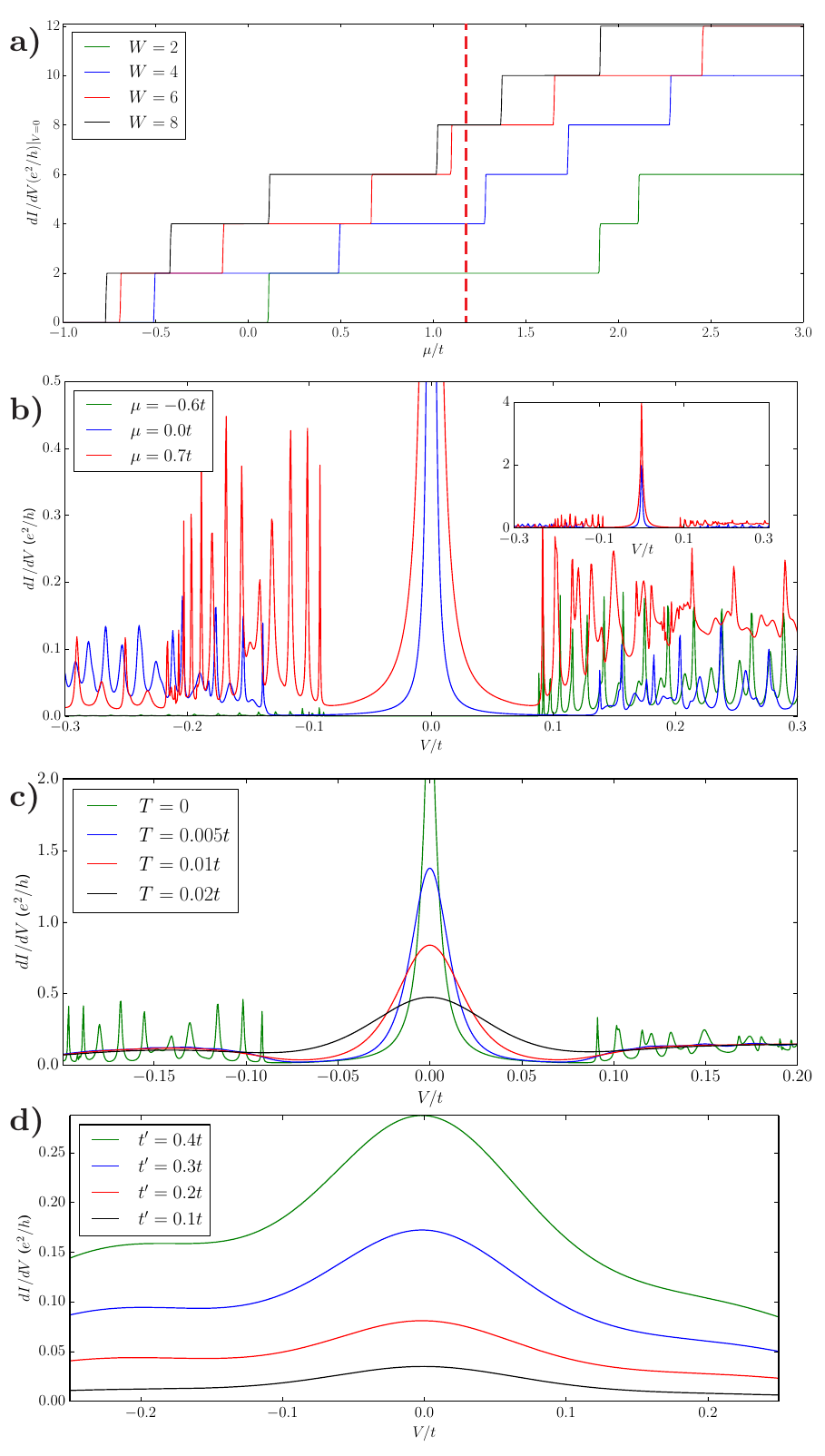}
\caption{(Color online)
(a) The zero-bias differential conductance as a function of chemical potential $\mu$ for various wire widths $W$ using a topological gap $\Delta=0.1t$ and a nanowire length of $L=200$ sites.
Since each Majorana modes contributes a factor of $2 e^2/h$ to the zero-bias signal, this measurement directly probes the number of Majorana states present.
For a given geometry the maximum possible conductance is $G_{max}=4  W  e^2/h$ (not shown here).
(b) Representative $dI/dV$ curves from parameter regimes with an integer topological invariant $|W| = (0,1,2)$ are given by the green, blue and red curves respectively.
Inset shows the quantized peak height for the blue ($2 e^2/h$) and red ($4 e^2/h$) curves.
(c) Finite temperature thermally broadens the zero-bias conductance peak width, thereby reducing the peak height to well below the quantized value of $2 e^2/h$.
(d) Weak STM - FM nanowire coupling, i.e. small $t'$, in conjunction with finite temperature ($T=0.05 t$) further reduces the peak height.
Note that the abscissa corresponds to an energy range much larger than the topological gap given by $\Delta=0.1t$.}
\label{fig:dIdV}
\end{figure}

We know from the topological properties discussion in Sec.~\ref{sec:topo} that MBS appear as soon as any FM bands, within a normal state picture, become occupied.
Setting $V=0$ and using Eq.~\ref{eq:dIdV} we see a peak in the zero-bias conductance, quantized in units of $2 e^2/h$, abruptly appearing at the critical value of the chemical potential when the first band becomes occupied ($\mu \approx -0.7t$) as shown in panel (a) of Fig.~\ref{fig:dIdV}.
As $\mu$ increases, higher sub-bands are filled while the Majorana occupancy increases, and each MBS contributes its own factor of $2 e^2/h$ to the total zero-bias differential conductance.
The zero-bias peak is a direct probe of the Majorana occupancy as the plot for $W=6$ in Fig.~\ref{fig:dIdV} (a) clearly mirrors the the zero-energy excitation spectrum given in Fig.~\ref{fig:Bands}.
In realistic experiments, because of disorder and electrical contact complications, it is difficult to increase the chemical potential uniformly across an entire sample in order to induce a topological phase transition.
It is for this reason that we instead propose manipulating the system width, i.e. tightly packing parallel magnetic atomic chains, as an experimental test of the chiral topological state.
Fig.~\ref{fig:dIdV} panel (a) illustrates the zero-bias signal behavior for various values of $W$.
Samples with different widths are expected to have a similar chemical potential, but the strength of the zero-bias peak at that $\mu$ should increase (red dashed line for example)
as a function of $W$.
It is also important to note that while finite temperature effects generally suppress the ZBCP height (as seen below), this effect is uniform and transitioning from $W=2$ to $W=4$ at $\mu$ corresponding to the dashed line, would still double the zero-bias signal.
The observation of such jumps in the ZBCP height with increasing the number of wires or channels will be a strong indication that the ZBCP is indeed arising from the localized MBS in the ferromagnetic wires in the BDI class.

Typical differential conductance profiles over finite voltage range are presented in Fig.~\ref{fig:dIdV} panel (b).
The green, red and blue lines are $dI/dV$ profiles generated for $\mu$ values corresponding to topologically distinct phases indexed by an integer winding invariant $|W| = (0,1,2)$.
A superconducting gap devoid of subgap states ($V \leq .1 t$) is characteristic of the trivial regime (green curve) while a quantized zero-bias signal appears in the non-trivial regimes.

The conductance at finite temperature $T$ is given by 
\begin{equation}
\label{eq:thermal}
\frac{dI(V,T)}{dV}=\int^{\infty}_{-\infty} dV' \frac{dI(V',T=0)}{dV'} \frac{d}{dV} f(V,T),
\end{equation}
where $f(E,V,T)=(\exp \left[ (E-\mu -eV)/ T \right] + 1)^{-1}$ is the fermi function. 
In this paper all zero temperature results will be assumed to be smeared by an infinitesimal temperature $T=10^{-5} t$. 
The finite temperature is crucial to avoid anomalies that depend on exponentially small coupling between Majorana modes which must exist in any finite length system no matter how long the wire is. 
As seen from previous calculations \cite{Flensberg} the zero-bias conductance vanishes at strictly zero temperature even for a topological system. 
However, this anomaly reduces to the usual result of a quantized conductance at temperature $T$ larger than the exponentially small Majorana splitting energy, but smaller than the tunneling energy between the Majorana mode and the lead. 
Strictly speaking, the tunnel coupling $t'$ between the STM tip and the Fe nanowire is unknown in the experiment of Ref.~\citenum{Yazdani_14} except that it is known to be very small. 
On the other hand, the experimental temperature in Ref.~\citenum{Yazdani_14} is very high, $>1K$, so the condition $t'>T$ is probably not satisfied in Ref.~\citenum{Yazdani_14}. 
Fortunately, this does not cause any qualitative problem in the theoretical analyses where most of the experimental parameters, except for the temperature, are not precisely known.
Thermally smeared differential conductance curves are plotted in Fig.~\ref{fig:dIdV} panel (c) for various temperatures.
Similar to the zero-bias phenomena observed in recent semiconductor experiments, where the peaks are generally an order of magnitude smaller than $2e^2/h$\cite{Mourik,Deng,Weizman,Churchill,Finck}, thermal effects smear our zero-bias peaks to well below its quantized value as was already pointed out in Ref.~\citenum{Long-PRB}.
Furthermore, the very weak coupling between the STM and the ferromagnetic nanowire, i.e. $t' \ll t$, in conjunction with finite temperature further reduces the ZBCP height.
Choosing a temperature of $T=0.02 t$ we illustrate this phenomena in Fig.~\ref{fig:dIdV} panel (d).

%
\begin{figure}[t!]
\centering
\includegraphics[width=\columnwidth]{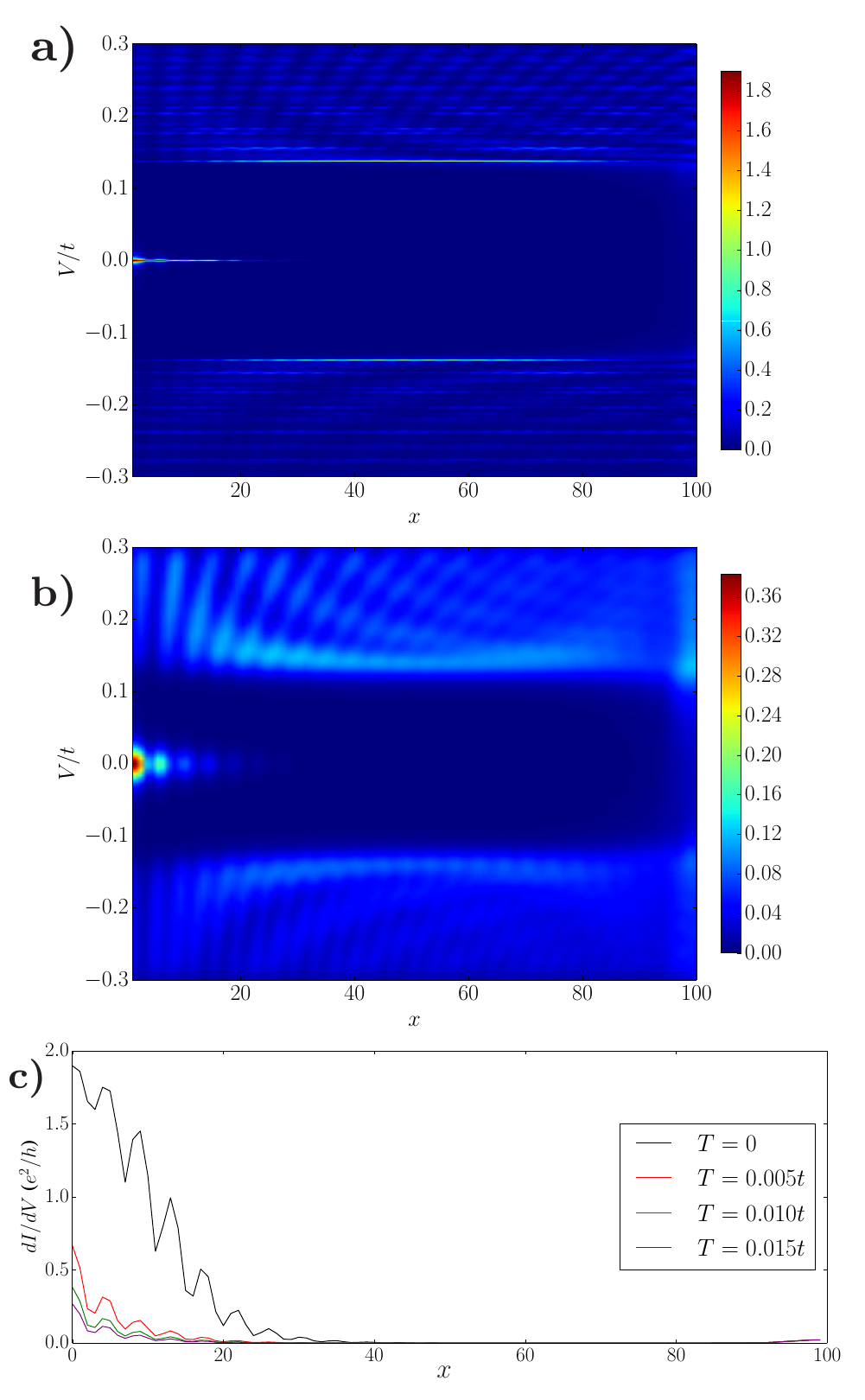}
\caption{(Color online)
Spatially resolved differential conductance profiles from the left edge to the middle of the nanowire. 
We use a topological pair potential $\Delta=0.1t$ and the length of the nanowire is $L=200$ sites.
Panel a) shows the zero temperature signal, while panel b) illustrates the effect of thermal smearing by a temperature smaller than the topological gap ($T=\Delta/10=0.01t$).
Note that for panels a,b) the energy is swept across the range $-3 \Delta$ to $3 \Delta$ and that the color scales in the top panels differ by an order of magnitude. 
Additionally panel c) highlights the spatial structure of the zero-bias signal for various temperatures.
See discussion below Eq.~\ref{eq:thermal} for a note regarding the $T=0$ result.}
\label{fig:spatial}
\end{figure}

By varying the STM coordinate $x$, we now simulate the tunneling spectra which would result from spatially sweeping the STM probe across the length of the sample, which has recently been experimentally achieved\cite{Yazdani_14}.
MBS are localized at each end of the wire, and we expect the ZBCP to vanish as the STM reaches the wire midpoint.
Fig.~\ref{fig:spatial} a (b) shows the zero (finite) temperature differential conductance spatial profile.
The signal due to tunneling into quasiparticle states above the superconducting gap remains approximately constant as the probe position varies, in contrast to the zero-bias signal which disappears in the bulk.
A zero-bias spatial profile displayed in Fig.~\ref{fig:spatial} (c) illustrates the exponential decay of the zero modes away from the edges as well as the end localization scaling with the characteristic length $\xi$ (see discussion in next section, Fig.~\ref{fig:psi}).
The features shown in Figs.~\ref{fig:dIdV} and \ref{fig:spatial} are generally consistent with the experimental findings in Ref.~\citenum{Yazdani_14}, providing some level of confidence that the experimentally observed ZBCP may indeed be arising from MBS-related physics (although the model parameters used in these figures are not realistic representations of the Fe/Pb system used in Ref.~\citenum{Yazdani_14}).
It is interesting to note that the very high temperature ($T \sim \Delta$) used in the experiment of Ref.~\citenum{Yazdani_14} along with the very small tunnel coupling ($t'$) to the STM tip not only leads to a very strong suppression of the ZBCP strength from its quantized Majorana value of $2e^2/h$, but also suppresses the range of $x$ values (indicating how far from the $x=0$ end point of the wire) over which the Majorana effectively `resides' as observed in the STM conductance measurement. 
The fact that the experimental Majorana observation seems to be localized near the wire end could just be a feature of the very high experimental temperatures. 
Thus, it is imperative that additional experimental data is obtained with higher values of $\Delta/T$ (either by increasing the effective topological superconducting gap or by lowing the experimental temperature) before one can reach a definitive conclusion regarding the existence or not of Majorana fermions in the experiment of Ref.~\citenum{Yazdani_14}.
We note that the experimental values of $\Delta/T$ in Ref.~\citenum{Yazdani_14} are much lower than those used in Figs.~\ref{fig:dIdV} and \ref{fig:spatial}, making it difficult, if not impossible, to reach any conclusion about the possible existence of Majorana fermions in the system.

In the next two sections, we provide a more detailed comparison between our numerical results and the experimental data of Ref.~\citenum{Yazdani_14}.

\section{Experimental Implications}

Having established generic features of our model, we now turn our attention to a comprehensive comparison with a recent experiment\cite{Yazdani_14} which shares many, \textit{but unfortunately not all}, features with our theoretical results.
Our focus here is mainly on comparing the qualitative phenomenological properties of the experimentally observed ZBCP and our theoretical results.
In addition to analyzing the height and width of the ZBCP as a function of temperature, wire length, and the STM tunnel barrier, we will also closely examine the spatial structure of the differential conductance profile, which can be directly calculated from our STM simulation (see Fig.~\ref{fig:spatial}).
To begin with, we first recapitulate the system parameters as quoted in Ref.~\citenum{Yazdani_14} and then set up our numerical parameters accordingly for comparison.
The ferromagnetic splitting is estimated to be $J=2.4\; eV $, which is much greater than the estimated hopping parameter $t = 1\;eV$ (which in turn is much larger than the superconducting gap $\sim 1\; meV$ in the substrate, thus allowing us to use the half-metallic ferromagnetic nanowire model for the theoretical description). 
Additionally, the superconducting gap in the underlying substrate is $\Delta_s = 1.36 \; meV$ while the induced $p$-wave gap is estimated to be $\Delta_p=100 \;  \mu eV$, although no direct nanowire superconducting gap with well-developed coherence peaks is visible at all in the experimental data presented \cite{Yazdani_14}.
Measurements were made on atomic chains between $5-15\; nm$ in length at a temperature $T=1.4K$ which corresponds to 100 $\mu eV$ in energy (roughly equal to the topological gap).
We mention here that these ferromagnetic nanowires are extremely short in length, containing only 10-50 Fe atoms only -- these wire lengths are by far the shortest lengths in the problem, being even shorter than the superconducting coherence length ($\sim 80 \; nm$) of Pb, the substrate superconducting material.
The topological coherence length in the Fe wire ($\sim 1000 \; nm$) is more than an order of magnitude larger than the length of the wires themselves. 
The estimate for the coherence length in the Fe wire obtained here assumes the induced topological superconducting gap to be $100-200\; \mu eV$ as provided in Ref.~\citenum{Yazdani_14}.
This experimentally used parameter regime is obviously a non-ideal regime for studying topological superconductivity since the lowest energy scale is the topological gap in the system, which is the same as the temperature of the system. 
Temperature would therefore be expected to suppress any signatures of the topological gap, including the Majorana zero-mode, which would merge with the bulk states. 

%
\begin{figure}[tb!]
\centering
\includegraphics[width=\columnwidth]{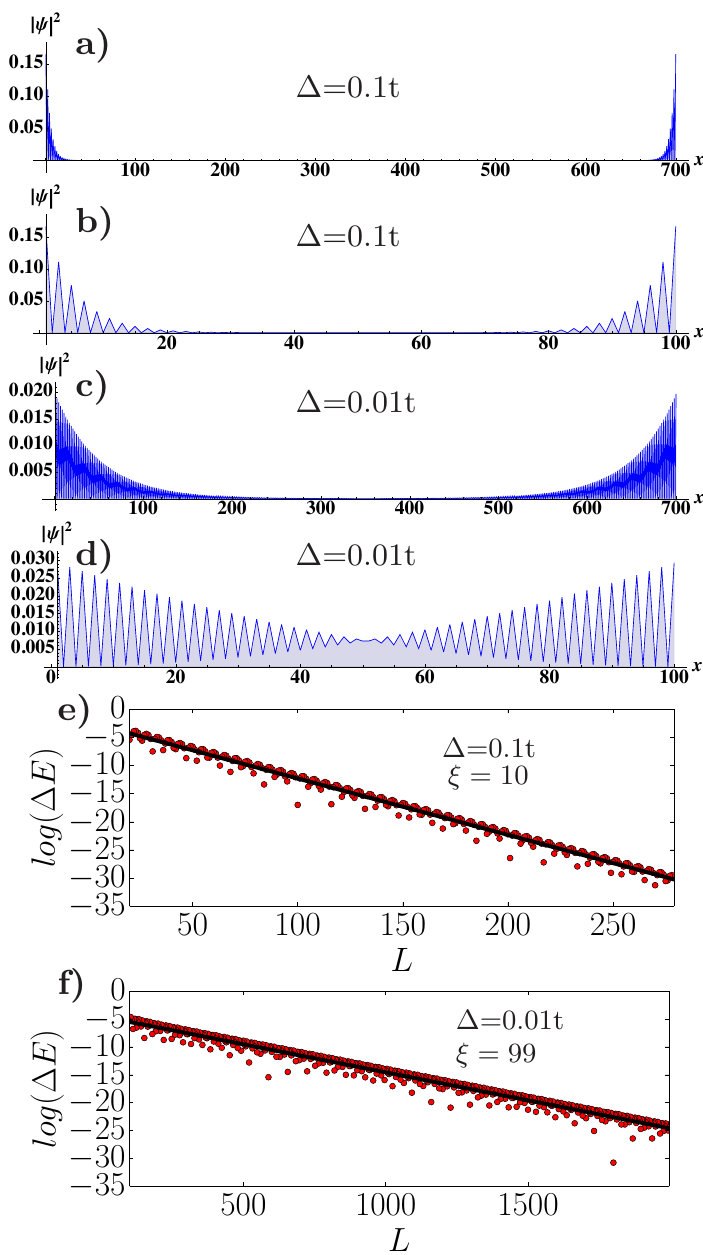}
\caption{(Color online) Majorana wavefunctions for small ($\Delta=0.01t$), and large ($\Delta=0.1t$), pairing regimes as calculated for short ($L=100$) and long $(L=700$) systems.
For both lengths considered in the large pairing regime, i.e. panels a,b), the Majorana decay length is much smaller than the system length ($\xi \ll L$) and the zero energy excitations are heavily localized to the FM wire endpoints.
Panels c,d) illustrate how reducing the pair potential to $\Delta=0.01t$ substantially increases the Majorana decay length (by an order of magnitude) and while a 700 site system still hosts Majorana states, considerable wavefunction overlap in the 100 site system hybridizes the end modes into non-zero energy, conventional delocalized quasiparticle states.
Panels e,f) show the logarithm of the energy splitting $\Delta E$ between the two Majorana modes.
As the wire length $L$ increases $\Delta E$ falls off exponentially.
Majorana decay lengths of of $\xi \sim 10$ sites, in the large pairing regime, and $\xi \sim 99$ sites, for the small pairing regime, are extracted from the black linear fits in panels e,f) respectively.}
\label{fig:psi}
\end{figure}
%

To compare our results with the experiment we introduce two parameter regimes, each characterized by different Majorana decay length scales, and then compare the results between the two regimes.
In the \textit{large pairing} regime we take the magnitude of the superconducting pair potential to be $\Delta_p=0.1t$ while in the \textit{small pairing} regime we use $\Delta_p=0.01t$. (To be clear, the system is always in a topological state and this is not to be confused with the weak/strong pairing regimes of Ref \citenum{Read-Green} which describe topologically non-trivial and trivial phases.)
In both cases we choose $\Delta_s=\Delta_p$ and simplify our notation by referring to this quantity simply as $\Delta$ (keeping in mind that $\Delta_p$ is responsible for the topological properties).
In both parameter sets we use $\mu=-.65t$, $W=2$ and $V_y=2t$.
We also typically choose very small values of $t'(\ll t)$ to simulate the very large tunnel barriers occurring at the STM tip contact with the nanowire.
(Very small values of $t'$ are essential for obtaining extremely weak zero-bias signals for the Majorana modes as observed experimentally.)

In a finite system, the localized MBS wavefunctions exponentially decay into the bulk with the characteristic superconducting coherence length $\xi \propto v_F/\Delta$, thus acquiring a finite energy due to wavefunction overlap from the two end MBS on two sides (true zero modes only occur in the $L  \ra \infty$ limit).
Fig.~\ref{fig:psi} panels (a,b) show the Majorana amplitude $|\Psi|^2$ on systems composed of $L =700,100$ sites in the large pairing regime.
The Majorana decay length is clearly much shorter than the wire length for both cases, so zero-energy Majorana bound states are localized at each end of the wire and with $|\Psi|^2$ being negligible near the midpoint.
Panels (c,d) illustrate the wavefunction amplitude in the small pairing regime, in which the Majorana decay length is comparable to the system size for the $L=100$ case.
The end modes appear unaffected in the $L=700$ wire, however, the wavefunctions for Majorana modes bound to opposite ends of the wire overlap significantly in the $L=100$ case, and as we discuss later, this has important ramifications for the zero-bias signal.
Note that the small gap used in the small pairing regime ($\Delta=0.01t$) is closer to the experimentally quoted parameters which would indicate a minuscule value of $\Delta=10^{-4}t$ (since the experimental system has $t \sim 1 eV$ and $\Delta \sim 100 \mu eV$).
We do not use even smaller $\Delta$ due to the prohibitive computational resources which would be required; however, the physics is generic and the topological pair potentials we have chosen are already sufficiently small to illustrate our point.
In fact, our theory is strongly over-emphasizing the topological aspects of the experimental systems-- all topological signatures will be weaker in the experiment compared with our results since the induced gap is smaller in Ref.~\citenum{Yazdani_14} than our chosen theoretical value.

$\Delta E$, the MBS splitting, can be directly captured using an effective Hamiltonian spanning the zero-energy Majorana subspace, $H_{eff} =  i(f/2) \gamma_L \gamma_R$ where $f \propto \ exp\( -L  / \xi \)$.
In Fig.~\ref{fig:psi} panels (e,f) we numerically calculate $\Delta E$ as a function of length in order to determine the decay length.
Plotted on a logarithmic scale, the red circles represent the raw data while the black linear regression has been fit to the data.
Taking $f = \ exp\( -L  / \xi \)$ we extract coherence (or equivalently, MBS localization) lengths of $\xi=10,100$ sites in the large and small pairing regimes respectively.
Note that the `beading' on top of the exponential decay is due to constructive and destructive interference between the MBS, and the length scale of these oscillations goes as $1/k_F$ \cite{Cheng}.

%
\begin{figure}[t!]
\centering
\includegraphics[width=\columnwidth]{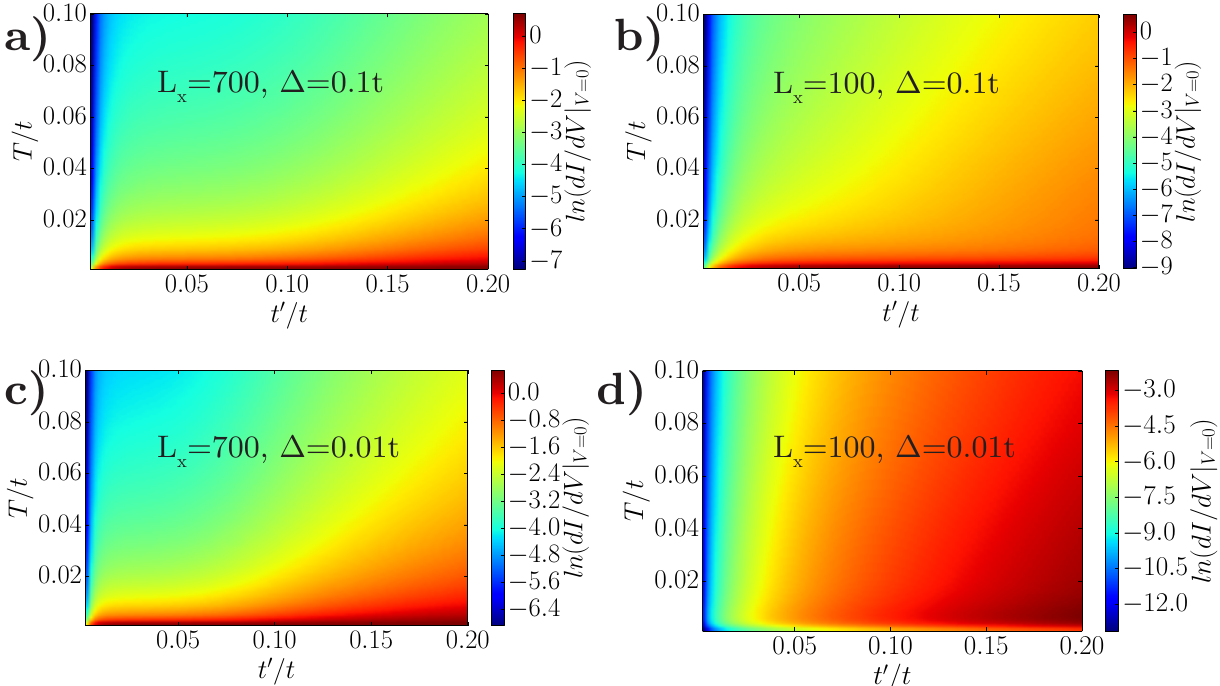}
\caption{(Color online) Zero-bias differential conductance peak height (panels correspond to same parameter values as used in Fig.~\ref{fig:psi} panels a-d), at one end of the wire ($x=0$) as a function of STM-FM coupling $t'$ and temperature $T$.
When the Majorana decay length is significantly shorter than $L $ (i.e. panels a-c) a quantized zero-bias signal of $2e^2/h$ (dark red on logarithmic color scale) is seen for zero temperature and $t'>0.03t$ (note that the STM-FM coupling in Ref.~\citenum{Yazdani_14} is expected to be smaller than this critical value of $t'$).
The quantized signal decays rapidly by introducing finite temperature or decreasing $t'$.
Panel d) No zero-bias signal is present near zero temperature (see below Eq. 9 for details) due to Majorana hybridization which splits the zero-bias peak into two separate finite bias signals.
The cause of this effect is finite temperature, which smears two finite-bias peaks together for an effective zero bias signal (see to Fig.~\ref{fig:L100_spatial} for details).}
\label{fig:coupling_temp}
\end{figure}

Next, we consider the roles of STM-FM coupling $t'$ and finite temperature $T$ in the quantitative suppression of the ZBCP strength in both parameter regimes.
As we have already noted in Fig.\ref{fig:dIdV} a small $t'$ reduces the ZBCP, which in conjunction with thermal smearing effects, significantly reduces the zero-bias signal.
As seen in Fig.~\ref{fig:coupling_temp} panels (a,b), 
the $T=0$ zero-bias signal in the large pair potential regime, where both wire lengths support MBS, saturates to the quantized value of $2e^2/h$ as $t'$ becomes large.
Increasing temperature or decreasing $t'$ both monotonically reduces the magnitude of the zero-bias signal, and similar behavior is found in panel (d) (long wire in small pairing regime), which also hosts well defined Majorana excitations.
Interestingly, as seen in panel (b), the short wire in the small pairing regime, i.e. one in which MBS have hybridized due to the small wire length ($\xi \sim L$), displays a finite temperature zero-bias signal comparable in magnitude to the finite temperature signal seen in the other panels.
Moving upward from the zero-temperature x-axis towards higher temperature, this signal grows until some critical value, after which the zero-bias signal decays like in the other panels.
As discussed later in this section, the source of this unusual zero-bias peak behavior increasing with temperature is thermal smearing between a pair of split Majorana states near zero energy.

%
\begin{figure}[t!]
\centering
\includegraphics[width=\columnwidth]{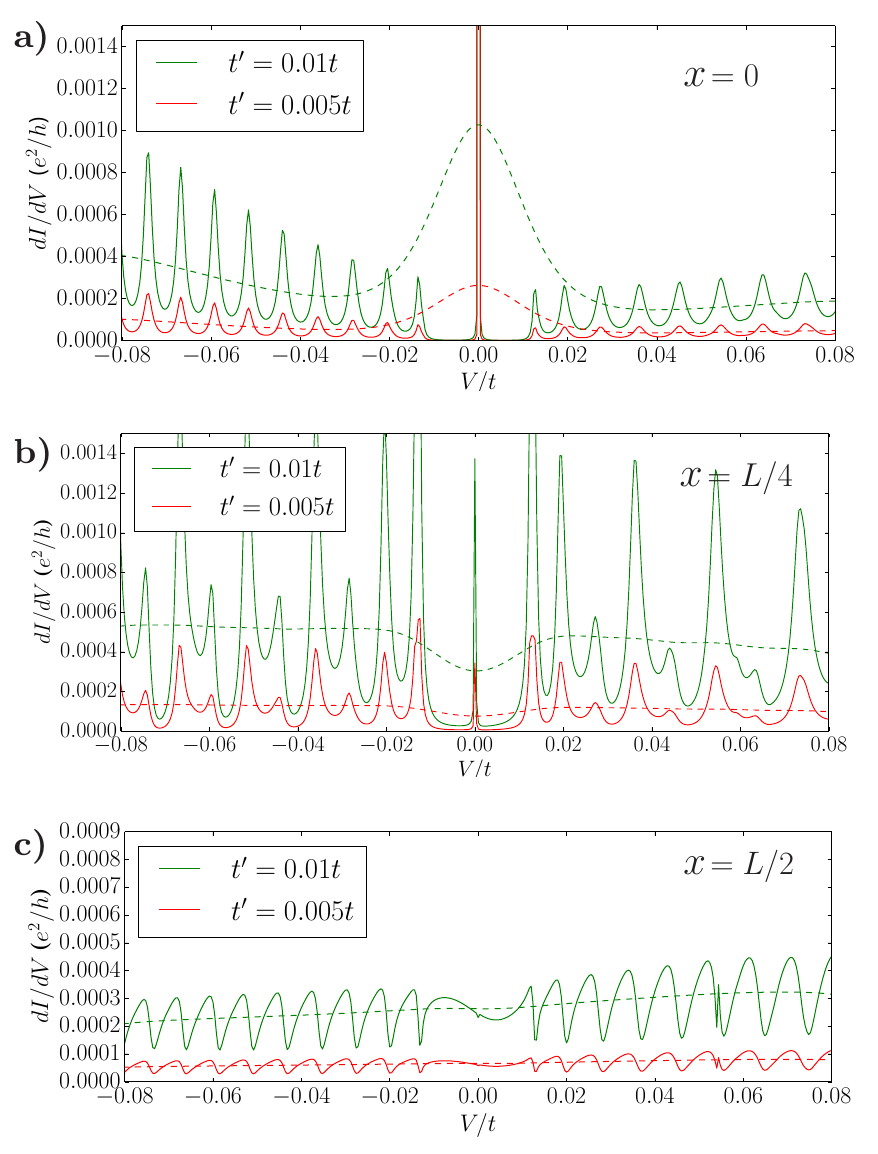}
\caption{(Color online)
Differential conductance calculated using an STM coordinate $x=0,L /4,,L /2$ for a long wire $L=700$ in the small pairing regime ($\Delta=0.01t$) using an STM-FM coupling $t'=0.01t$.
Solid and dashed lines are at temperature $T=0,0.01t$ respectively, and green (red) lines correspond to a STM-FM coupling of $t'=0.01t \; (0.005t)$.
Panel a) shows a clear zero-bias signature which is visible at the wire endpoints for both zero and finite temperature.
The green dashed line ($T=\Delta$) displays a peak height ($10^-3 e^2/h \sim 40 nS$) and width ($\sim \Delta$) which are comparable to the experimentally reported values.
Panels b,c) The zero temperature Majorana peak decays as the STM moves into the FM nanowire bulk.
The peak completely vanishes at the midpoint, and is not visible at zero or finite temperature signal.
Note that the energy scale for the abscissa is much larger than the topological gap in these figures.}
\label{fig:L700}
\end{figure}
\begin{figure}[t!]
\centering
\includegraphics[width=\columnwidth]{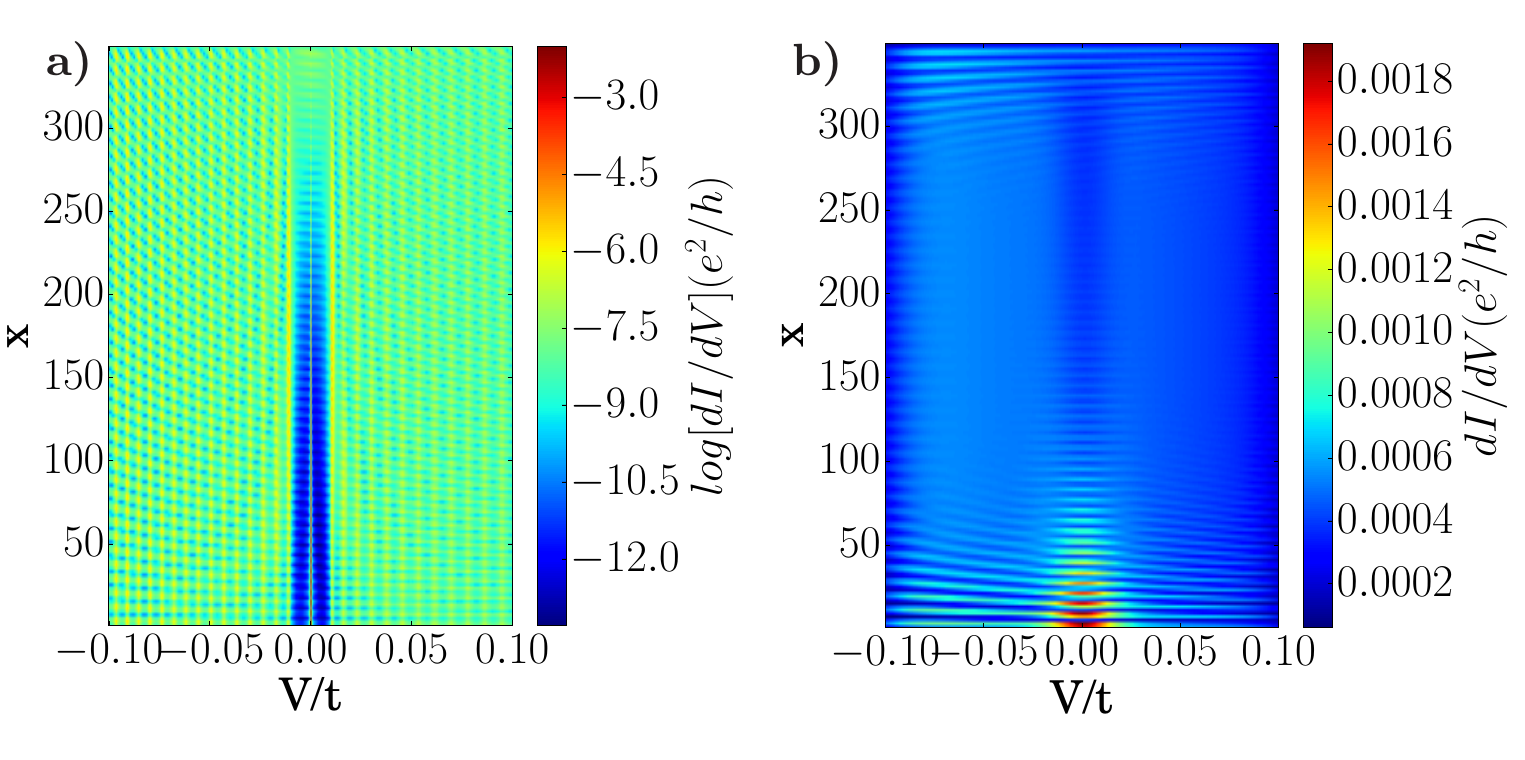}
\caption{(Color online)
Differential conductance spatial profile as a function of the STM coordinate $x$ for a long wire $L=700$ in the small pairing regime ($\Delta=0.01t$) using an STM-FM coupling $t'=0.01t$.
Panels a,b) correspond to temperature $T=0,0.01t$ respectively, and we present the zero temperature data on a logarithmic scale for improved visibility.
At zero temperature the spatially localized Majorana mode (the localization length appears longer on a logarithmic scale) resides within a well defined superconducting gap which is not visible at finite temperature due to thermal smearing effects.
Note that the true spatial extent of the Majorana mode is revealed by the finite temperature result which is plotted on a linear scale and that the energy scale for the abscissa is much larger than the topological gap in these plots.}
\label{fig:L700_spatial}
\end{figure}

Having established that a strongly suppressed ZBCP is a generic feature of the experimental parameter regime (i.e. small $t'$, large $T$, and small topological gap), observable with or without the existence of zero energy Majorana excitations, we now analyze the spatial profile of the ZBCP, which can in principle be used to distinguish between a signal originating from zero energy or finite-energy split quasi-MBS.
Focusing on the small pairing parameter regime first, i.e., the parameters which are closer to those reported in Ref.~\citenum{Yazdani_14}, we plot the differential conductance measured at three STM positions $x = 0 ,L/4 , L/2$ along a wire of length 700 sites (see Fig.~\ref{fig:L700}).
In panel (a) the conductance is measured from the wire endpoint (i.e. $x=0$) and we observe that finite bias quasiparticle states are separated from the Majorana signal by a gap which is comparable in magnitude to the pair potential (recall, $\Delta=0.01t$).
Note the zero-bias Majorana signal (green and red solid lines are almost completely superimposed and therefore not discernible) is delta function shaped as a consequence of the small coupling parameter $t'$.
The dashed lines indicate the signal at a finite temperature of $T=\Delta=0.01t$, which is the case in Ref.~\citenum{Yazdani_14}.
The thermally broadened peak height for the green dashed line is $10^{-3} e^2/h \approx 40 nS$ which is comparable to that reported in the experiment.
Additionally, the peak width at half maxima is $\Delta$, which is also consistent with experimental results.
Panel (b) shows how by moving the STM tip into the bulk of the wire ($x=L/4$) the zero-bias signal drastically falls off, to the point where it is of the same order of magnitude as the conventional background thermal quasiparticle signal.
Due to the large separation between the zero- and finite-bias signals, a valley, centered around $V=0$, appears in the thermally broadened conductance profile (dashed line).
Lastly, in panel (c) we see that, as expected, the Majorana peak is completely absent at the wire midpoint $x=L/2$.
All these features appear to be qualitatively consistent with the experimental data reported in Ref.~\citenum{Yazdani_14}.

A detailed spatial profile of the differential conductance should reveal the highly localized Majorana wavefunction from Fig.~\ref{fig:psi}.
In order to numerically reveal the localized nature of these Majorana wavefunctions, we smoothly vary the STM tip position $x$ and plot the zero temperature differential conductance at each point in Fig.~\ref{fig:L700_spatial} panel (a).
Along most of the wire, the spatially resolved $dI/dV$ indicates a well formed superconducting gap separating the single Majorana peak at zero energy from the finite energy quasiparticles.
Note that, in this plot, the spatial extension of the Majorana wavefunction is exaggerated due to the logarithmic scale which has been used to increase the visibility of the data.
Panel (b) shows the spatially resolved conductance at finite temperature ($T=\Delta$), which reveals the true spatial extent of the Majorana mode.
We pause to note that while the model parameters used here are similar those quoted in the experiment, the localization length seen in panel (b) is significantly larger than reported in Ref.~\citenum{Yazdani_14}.

%
\begin{figure}[tb!]
\centering
\includegraphics[width=\columnwidth]{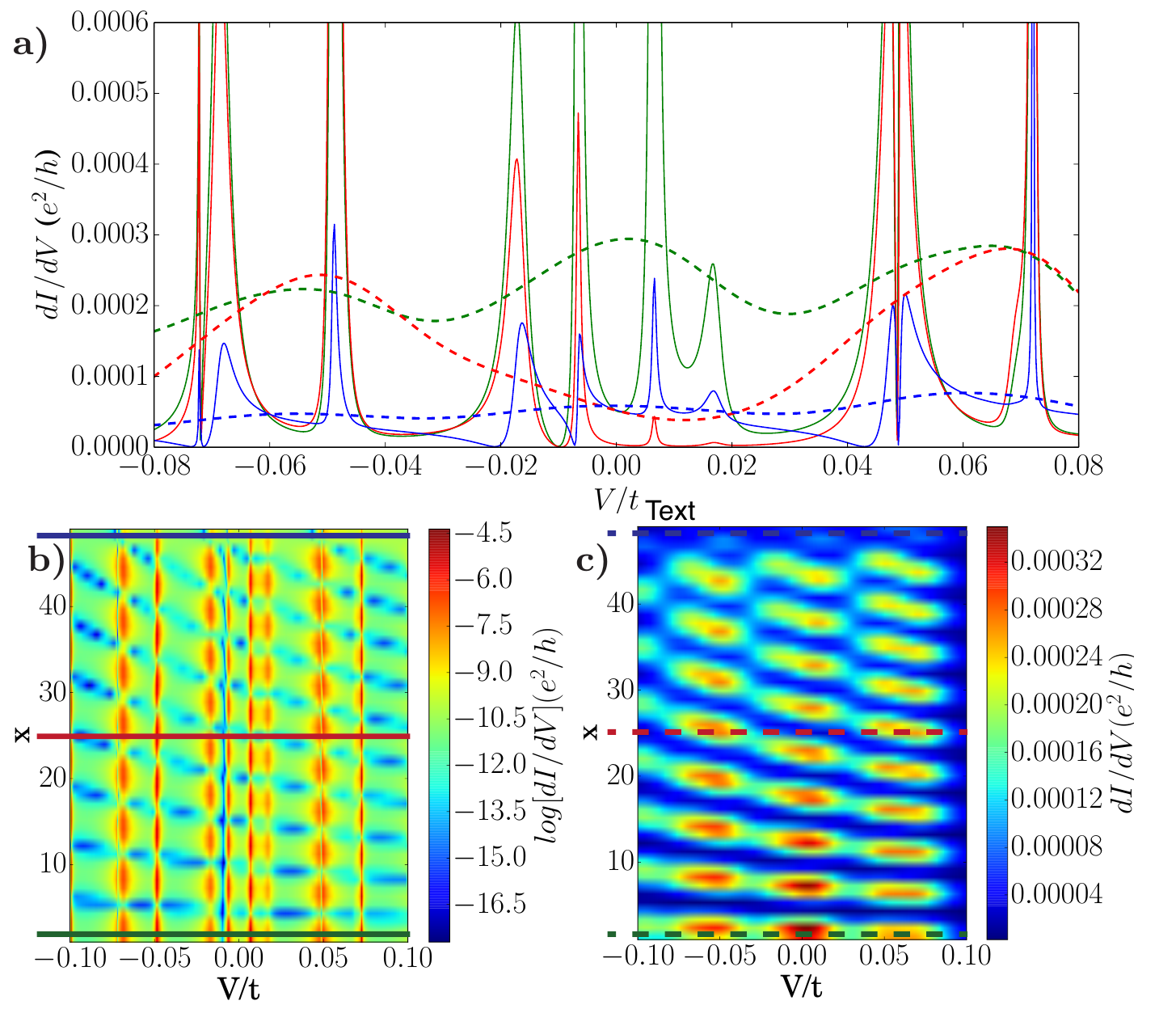}
\caption{(Color online)
When the ferromagnetic nanowire is short (i.e. $L  \sim \xi$ ) Majorana bounds states at opposite ends hybridize as their wavefunctions overlap significantly in the bulk.
Panel a) Using the same parameters from the Fig.~\ref{fig:L700_spatial} except $L =100$, we see that the zero-bias Majorana signature splits into two distinct peaks centered near $V=\pm 0.01t$.
The green, red, and blue curves correspond to the differential conductance calculated at $x=0, L  / 4, L  / 2$ respectively, and the solid (dashed) lines denote the zero (finite $T=\Delta=0.01t$ ) temperature.
Panels b,c) Spatially resolved differential conductance along the longitudinal axis for zero (on a log scale) and finite temperature (linear scale).
In panel c) we see that the split Majorana conductance peaks are indiscernible at finite temperature due to thermal smearing (signal at $V \sim 0$).
This finite temperature signal is flanked on both sides by broadened conventional quasiparticle peaks.
Solid and dashed lines are used to indicate the STM position for the $dI/dV$ curves presented in panel a).
Note that the energy scale for the abscissa is much larger than the topological gap ($\Delta=0.01t$) in these plots.}
\label{fig:L100_spatial}
\end{figure}

Since small pair potentials and short wire lengths are quoted in Ref.~\citenum{Yazdani_14}, we now investigate the small pair potential regime ($\Delta=0.01t,\xi=99$) on the $L=100$ system, i.e. parameters which should be most applicable to the experiment.
Remember that for these parameters the Majorana modes generally hybridize (see Fig.~\ref{fig:psi}) and therefore the Majorana splitting should be visible.
Using an STM-FM coupling strength $t'=0.01t$, the green, red and blue solid lines in Fig.~\ref{fig:L100_spatial} panel (a) show the differential conductance calculated at STM positions $x=0,L/4,L/2$.
We immediately note that, due to finite size effects, the zero-bias signal has split into two peaks centered around $V=0$, with an estimated energy splitting comparable to the gap energy ($\Delta E=0.01t$).
Also note, that while the tunneling signal into these two finite energy quasiparticle states may be the largest at $x=0$, the signal persists well into the bulk of the wire ($x=L/4,L/2$).
At finite temperature $T=\Delta$, the splitting between the peaks is no longer visually resolvable since temperature has thermally broadened the signal across a range greater than the original Majorana splitting $\Delta E$ (dashed green line).
Panels (b,c) illustrate the differential conductance spatial profile for temperatures $T=0,0.01t$, where again we have presented the zero temperature data on a logarithmic plot.
In these bottom panels we see that the split quasi-Majorana modes are spatially extended across the entire length of the wire.
Green, red and dark blue solid and dashed lines superimposed on the spatial profiles correspond to the $dI/dV$ curves presented in panel (a).
Thus, in the context of a small pairing potential present in a short wire, we see a finite temperate zero-bias signal extending into the bulk of the wire, with no measurable decay.
For reasons which remain unclear at this stage, this theoretically expected splitting of the Majorana mode and the associated spatial delocalization of quasi-Majorana modes are again not observed in the experiment, but, see the next section for a possible resolution of this puzzle.

\begin{figure}
\includegraphics[width=\columnwidth]{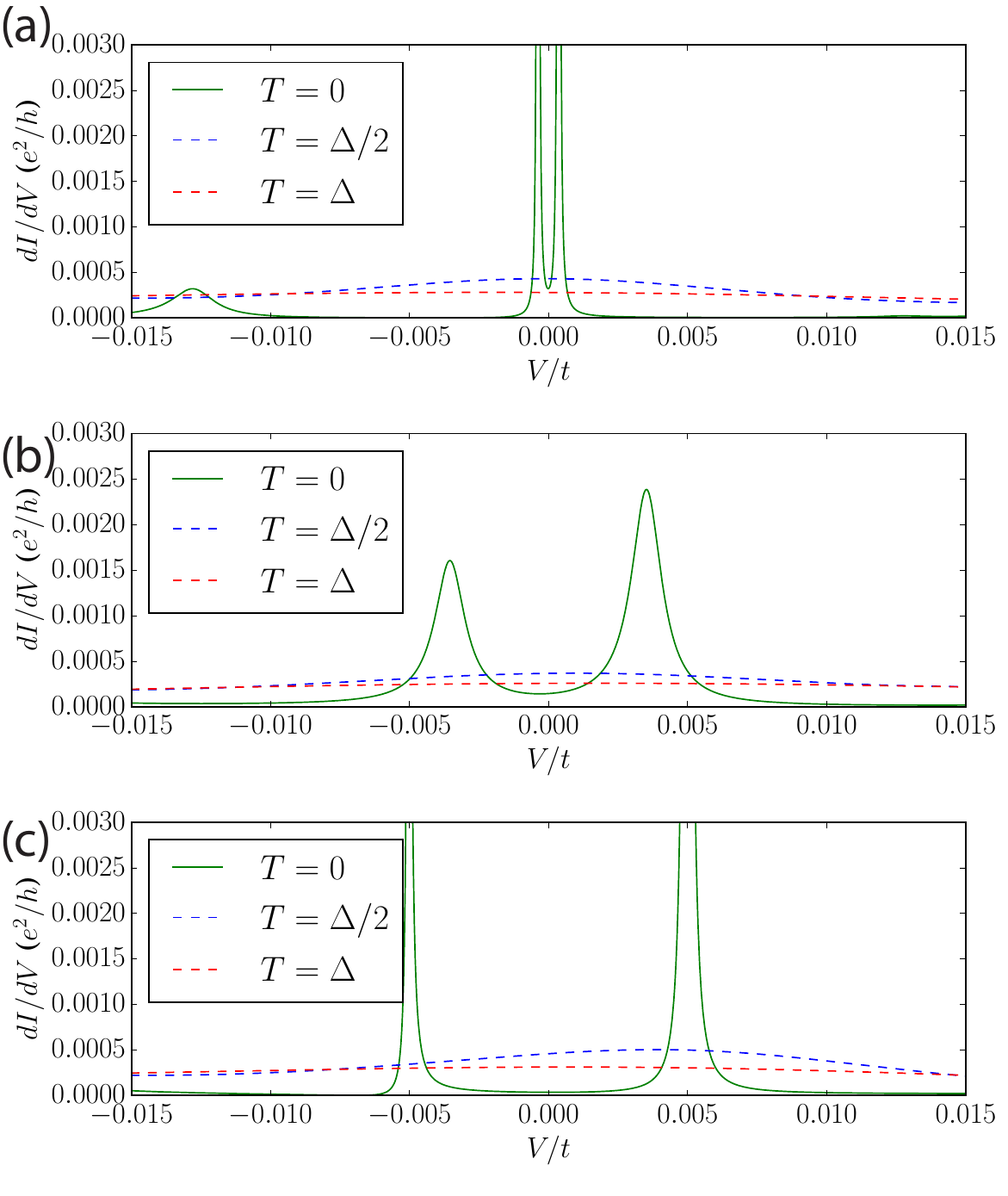} 
\caption{Evolution of the splitting of the Majorana conductance peak, plotted across $V=(-1.5 \Delta,1.5 \Delta)$, for systems of length $L=300 (a), 200 (b),100 (c)$. 
We have taken the pairing potential to be $\Delta=0.01t$ and have calculated the conductance from the end of the wire (STM positioned at $x=0$ with an STM nanowire coupling $t'=0.01t$).
The solid (dashed) curves show the differential conductance at zero (finite) temperature.
Because the three lengths used are comparable to the superconducting coherence length ($\xi=100$ sites), the Majorana modes hybridize into finite energy quasiparticles with energies $\pm \Delta E$.
This hybridization splits the ZBCP into two peaks which are indiscernible at finite temperature due to thermal smearing effects.}
\label{fig:end_splitting_SPR}
\end{figure}
\begin{figure}
\includegraphics[width=\columnwidth]{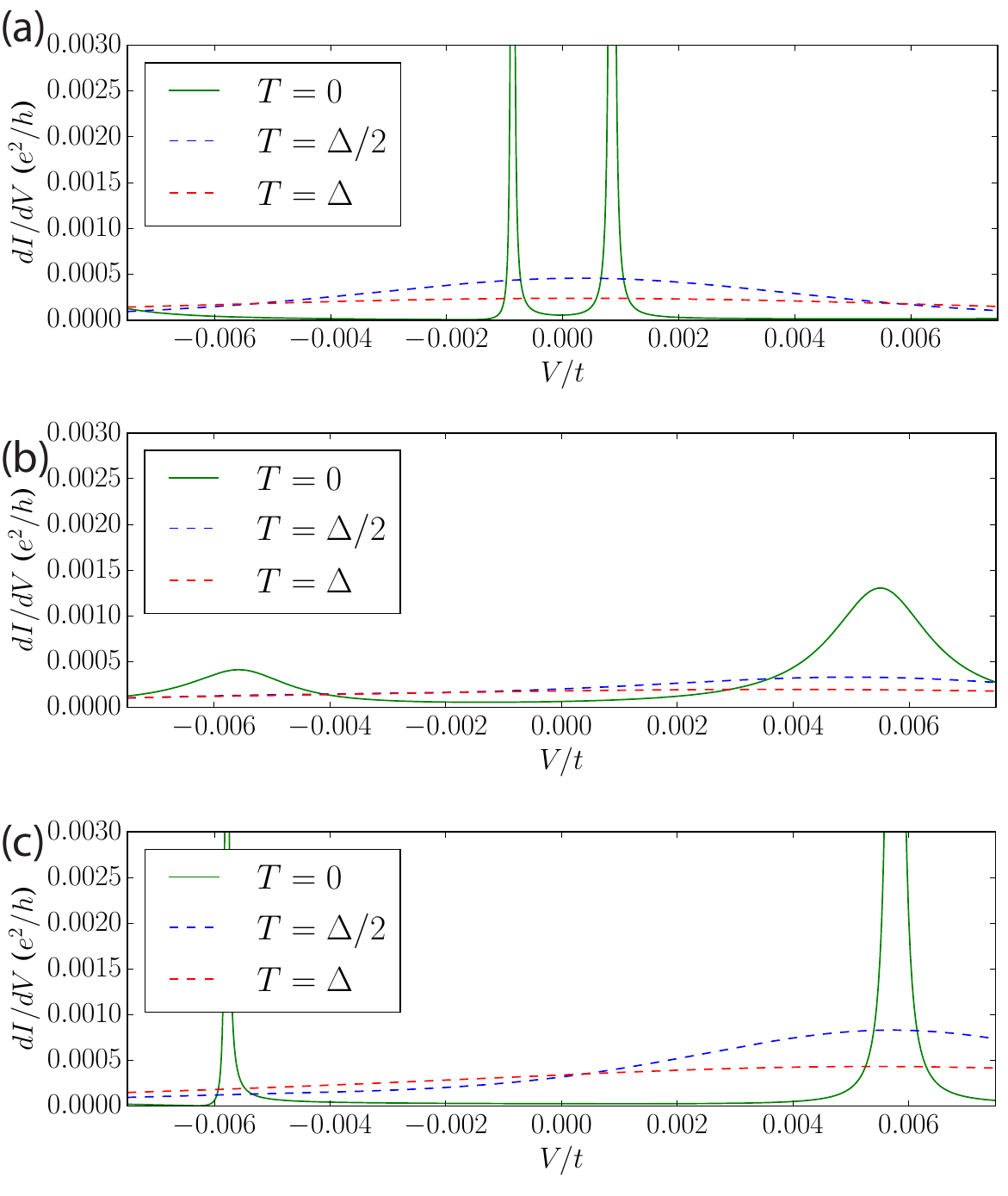} 
\caption{Evolution of the splitting of the Majorana conductance peak, plotted across $V=(-1.5 \Delta,1.5 \Delta)$, for systems of length $L=300 (a), 200 (b),100 (c)$.
Here we use a smaller pairing potential of $\Delta=0.005t$ and have again calculated the conductance from the end of the wire (STM positioned at $x=0$ with an STM nanowire coupling $t'=0.01t$).
The solid (dashed) curves show the differential conductance at zero (finite) temperature.
Because the three lengths used are comparable to the superconducting coherence length ($\xi=200$ sites), the Majorana modes hybridize into finite energy quasiparticles with energies $\pm \Delta E$.
Again, this hybridization splits the ZBCP into two peaks (distinct at zero temperature) which are indiscernible at finite temperature due to thermal smearing effects.}
\label{fig:end_splitting_TPR}
\end{figure}
\begin{figure}
\includegraphics[width=\columnwidth]{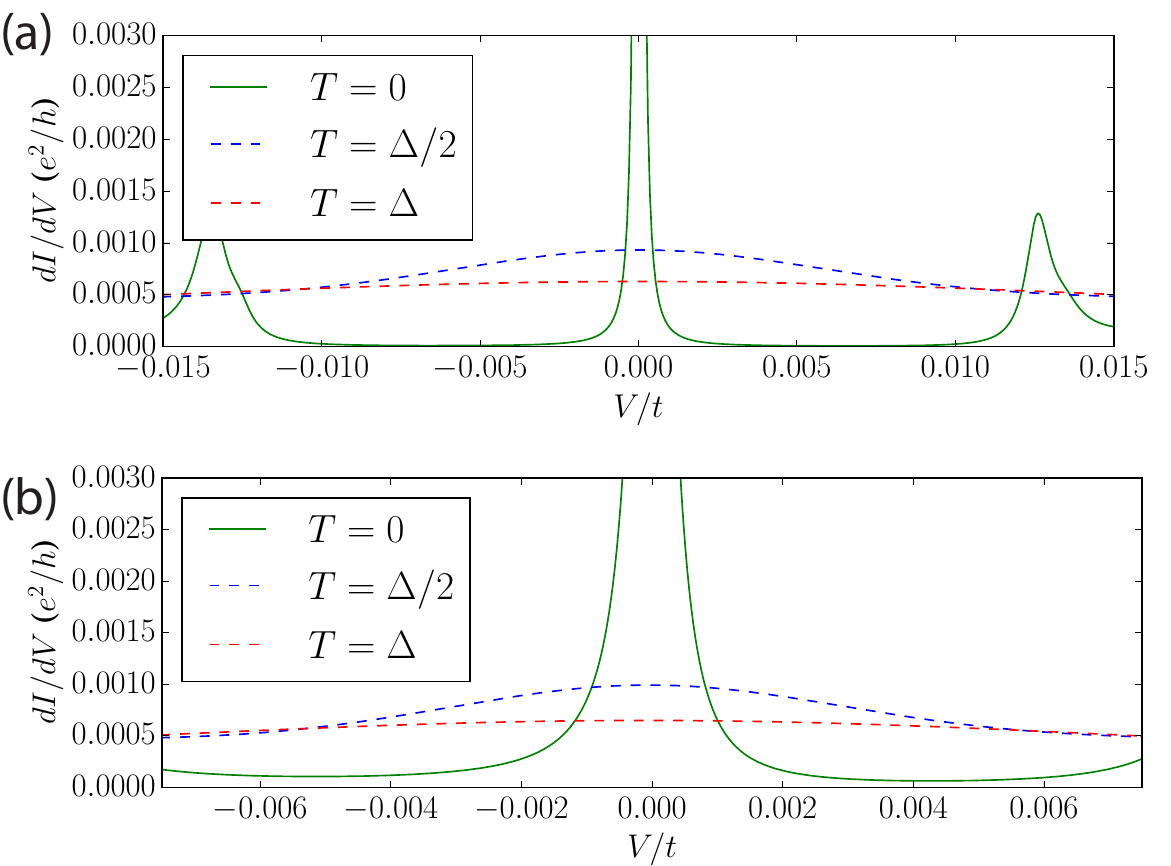} 
\caption{Differential conductance tunneling spectrum, plotted across $V=(-1.5 \Delta,1.5 \Delta)$, for systems of length $L=700$. 
The weakly coupled ($t'=0.01t$) STM is positioned at at $x=0$ and we use a pair potential of $\Delta =0.01t$ (corresponding coherence length $\xi=100$ sites) and $\Delta=0.005t$ ($\xi=200$ sites) in panels (a,b) respectively. 
The solid (dashed) curves show the differential conductance at zero (finite) temperature.
Since the system length is greater ($L \gg \xi $) than the superconducting coherence length no peak splitting is observed.}
\label{fig:long_no_splitting}
\end{figure}
%
\begin{figure*}
\includegraphics[width=14cm]{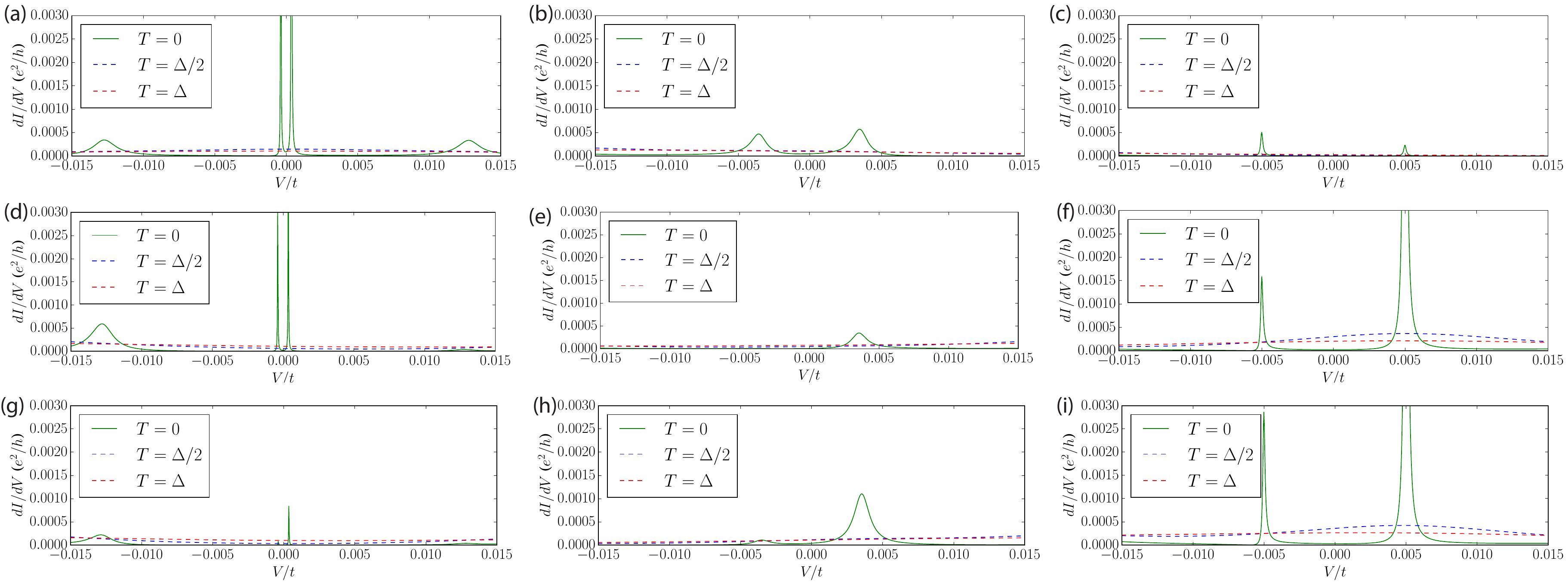}
\caption{Evolution of the splitting of the Majorana conductance peak, plotted across $V=(-1.5 \Delta,1.5 \Delta)$, for systems of length $L=300 (\text{left column}), 200 (\text{middle column}),100 (\text{right column})$. 
We have taken the pairing potential to be $\Delta=0.01t$ and the weakly coupled ($t'=0.01t$) STM is now positioned away from the system edge at $x=L/4$ for panels (a-c), at $x=L/4+2$ for panels (d-f), and at $x=L/4+3$ for panels (g-i).
The solid (dashed) curves show the differential conductance at zero (finite) temperature.
Because the three lengths used are comparable to the superconducting coherence length ($\xi=100$ sites), the Majorana modes hybridize into finite energy quasiparticles with energies $\pm \Delta E$.
This hybridization splits the ZBCP into two peaks which are indiscernible at finite temperature due to thermal smearing.}
\label{fig:L/4_splitting_SPR}
\end{figure*}
\begin{figure*}
\includegraphics[width=14cm]{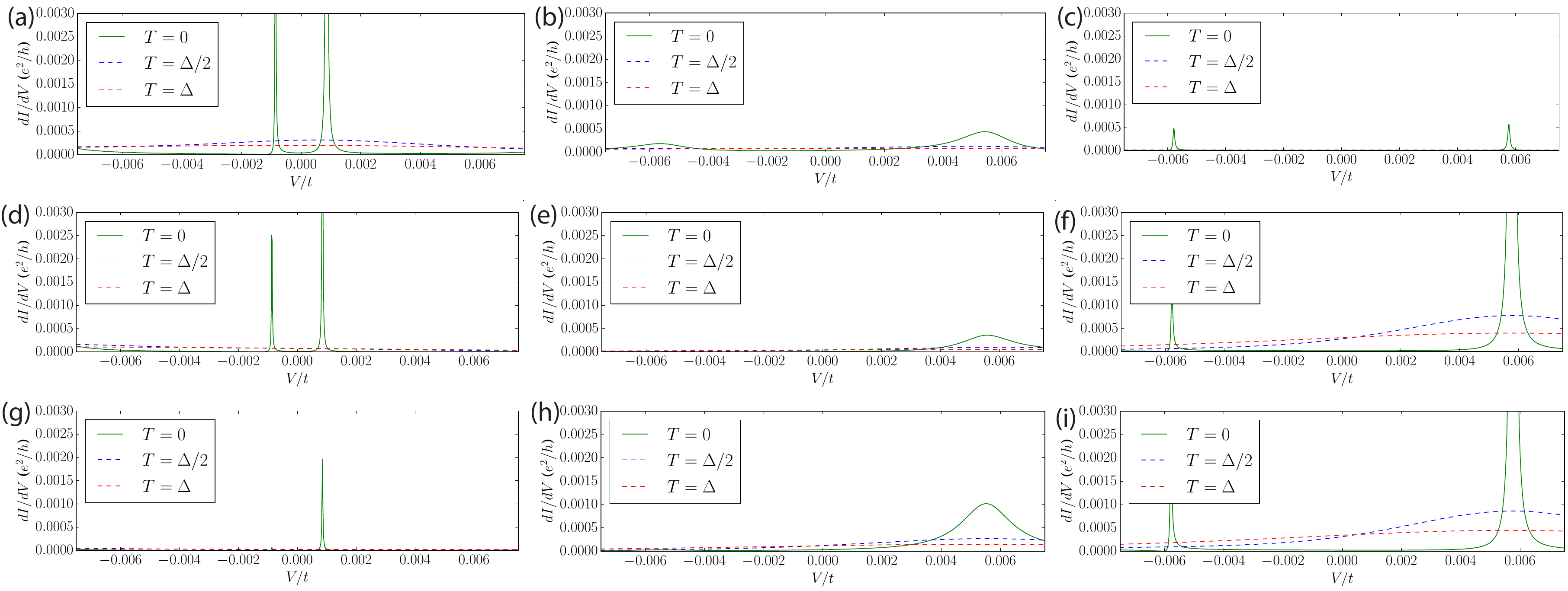}
\caption{Evolution of the splitting of the Majorana conductance peak, plotted across $V=(-1.5 \Delta,1.5 \Delta)$, for systems of length $L=300 (\text{left column}), 200 (\text{middle column}),100 (\text{right column})$. 
We have taken the pairing potential to be $\Delta=0.005t$ and the weakly coupled ($t'=0.01t$) STM positioned at $x=L/4$ for panels (a-c), at $x=L/4+2$ for panels (d-f), and at $x=L/4+3$ for panels (g-i).
The solid (dashed) curves show the differential conductance at zero (finite) temperature.
Because the three lengths used are comparable to the superconducting coherence length ($\xi=200$ sites), the Majorana modes hybridize into finite energy quasiparticles with energies $\pm \Delta E$.
This hybridization splits the ZBCP into two peaks which are indiscernible at finite temperature due to thermal smearing.}
\label{fig:L/4_splitting_TPR}
\end{figure*}
\begin{figure}
\includegraphics[width=\columnwidth]{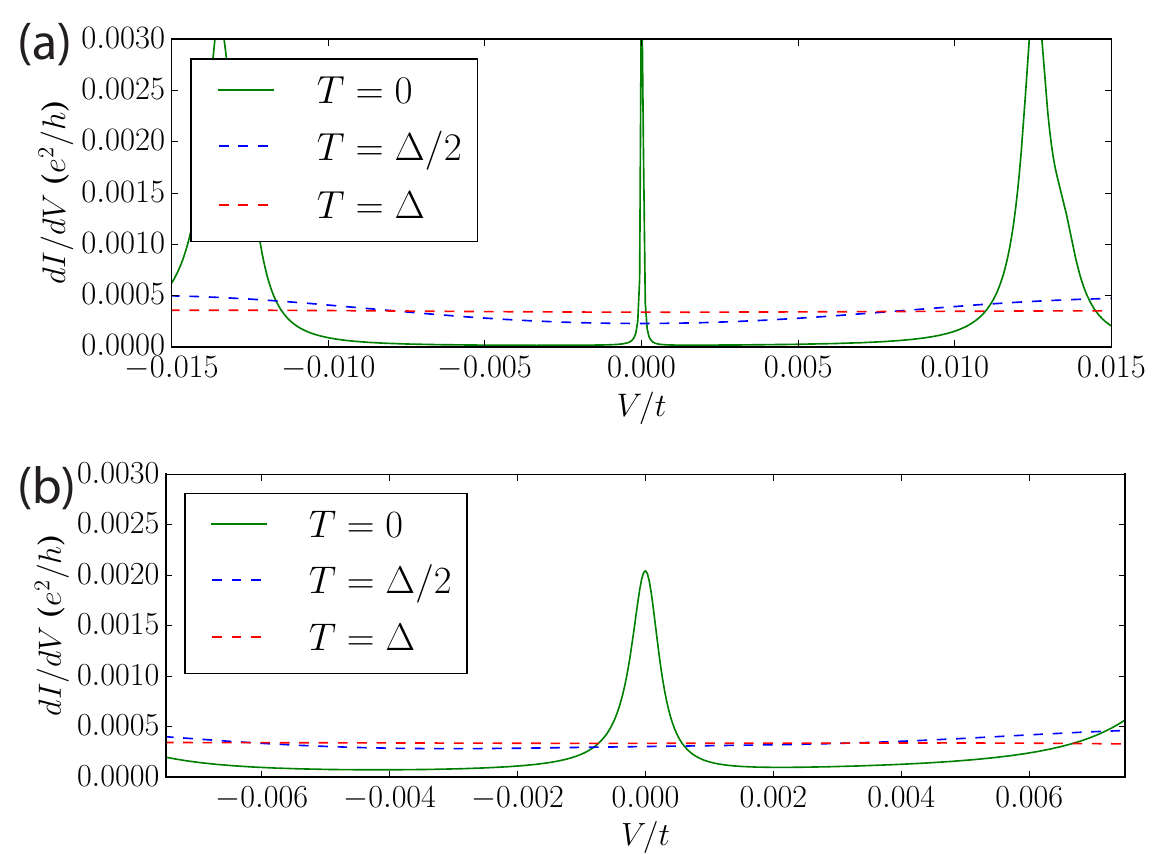} 
\caption{Differential conductance tunneling spectrum, plotted across $V=(-1.5 \Delta,1.5 \Delta)$, for systems of length $L=700$. 
The weakly coupled ($t'=0.01t$) STM is positioned at at $x=L/4$ and we use a pair potential of $\Delta =0.01t$ (corresponding coherence length $\xi=100$ sites) and $\Delta=0.005t$ ($\xi=200$ sites) in panels (a,b) respectively. 
The solid (dashed) curves show the differential conductance at zero (finite) temperature.
The system length is much greater ($L \gg \xi $) than the superconducting coherence length and no Majorana splitting is observed.}
\label{fig:long_no_splitting}
\end{figure}
%
%
\begin{figure*}
\includegraphics[width=14cm]{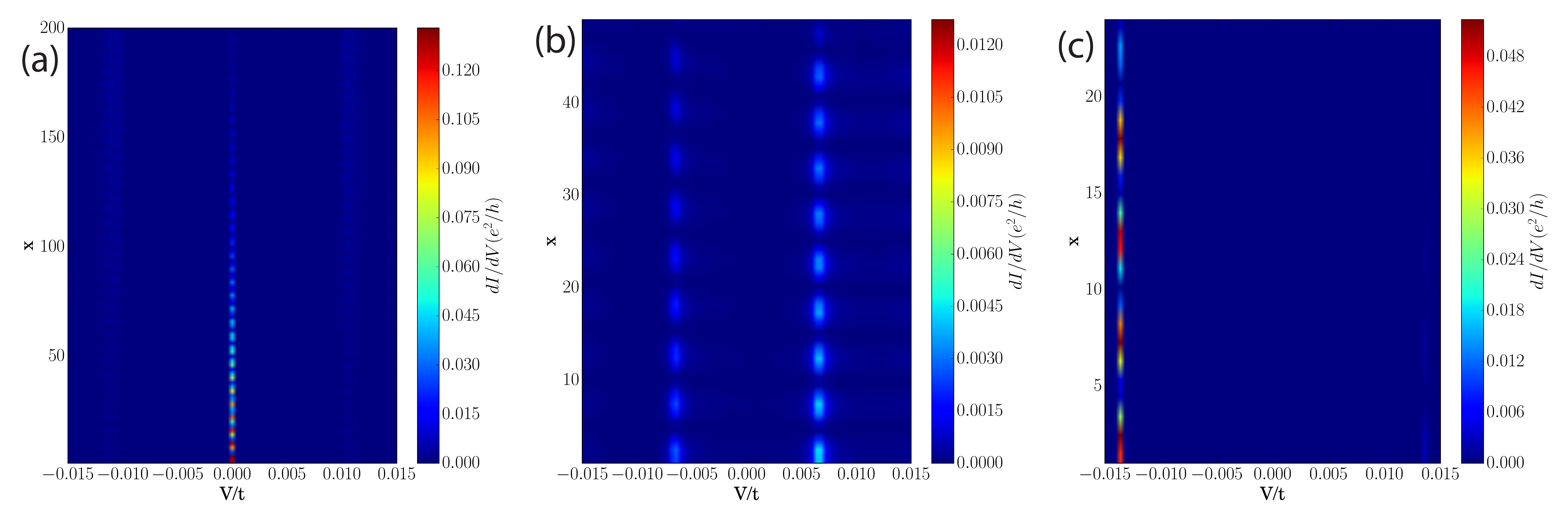} 
\caption{Spatially resolved differential conductance profiles for a systems of length $L=700,100,50$ in panels (a,b,c) respectively.
The differential conductance is calculated at zero temperature ($T=0$) with a topological gap of $\Delta=0.01t$ (the coherence length is $\xi=100$ sites) and a weak STM-nanowire coupling is $t'=0.01t$.
In panel (a) the nanowire is much longer than the coherence length which results in a zero-bias signal, which rapidly decays into the bulk over a range of $2\xi=100$ sites, due to the localization associated with a Majorana end mode. 
In contrast, the Majorana modes split to finite energy in panels (b,c) as a result of a short nanowire length ($\xi \gtrsim L$).}
\label{fig:0T_profile}
\end{figure*}
\begin{figure*}
\includegraphics[width=14cm]{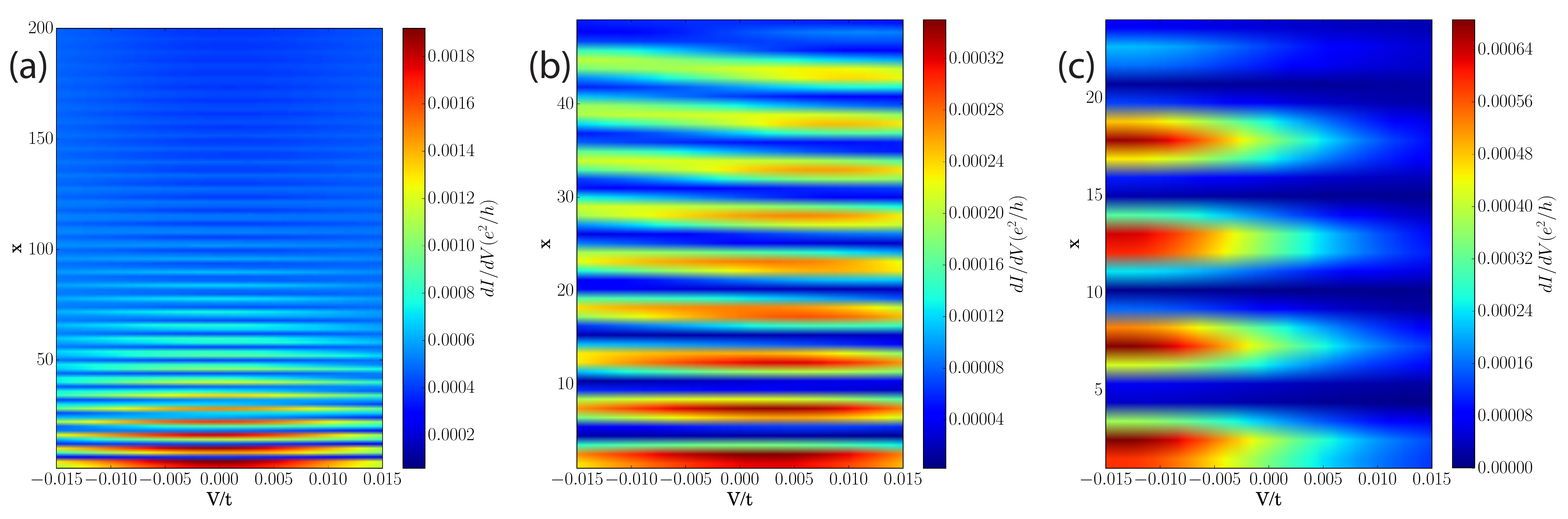} 
\caption{Spatially resolved differential conductance profiles for a systems of length $L=700,100,50$ in panels (a,b,c) respectively.
The calculation is performed at a finite temperature ($T=\Delta$), with a topological gap of $\Delta=0.01t$ ($\xi=100$ sites), and with a weak STM-nanowire coupling is $t'=0.01t$.
At this temperature the zero-bias Majorana peak is thermally broadened across the entire width of the topological gap but remains localized to the nanowire end.
The hybridized Majorana peaks, appearing in shorter nanowires as seen in panels (b,c), share the qualitative feature (with a true Majorana signal) that they are smeared across the entire gap at high temperature.}
\label{fig:finiteT_profile}
\end{figure*}
\begin{figure}
\begin{centering}
\includegraphics[width=\columnwidth]{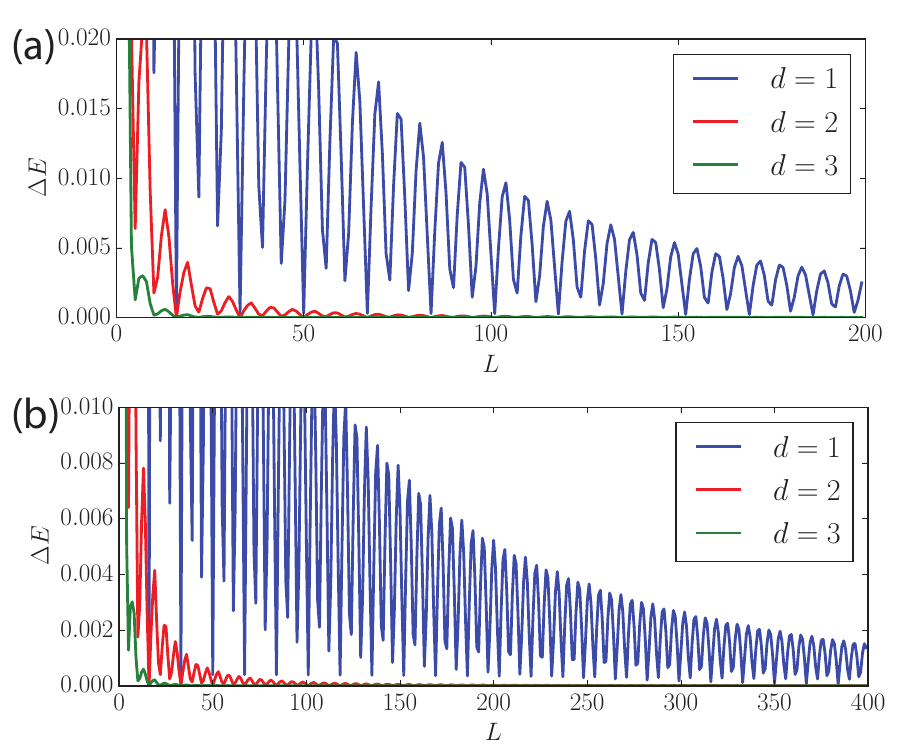} 
\caption{Majorana splitting energy $\Delta E$ (determined by exact diagonalization) as a function of the system length $L$, in $d=1,2,3$ spatial dimensions.
In panel (a) we use $\Delta = 0.01t$ (i.e. the small pairing regime with a coherence length of $\xi=100$ sites) while in panel (b) we choose an even smaller value of $\Delta =0.005t$ (with a coherence length of $\xi=200$ sites).
The splitting in $d=2,3$, (red and green lines), differs from the one-dimensional result by a factor of $1/L$ and $1/L^2$ respectively, yielding a more pronounced decay as length increases.}
\label{fig:2D3DSplitting}
\end{centering}
\end{figure}
\begin{figure*}
\includegraphics[width=14cm]{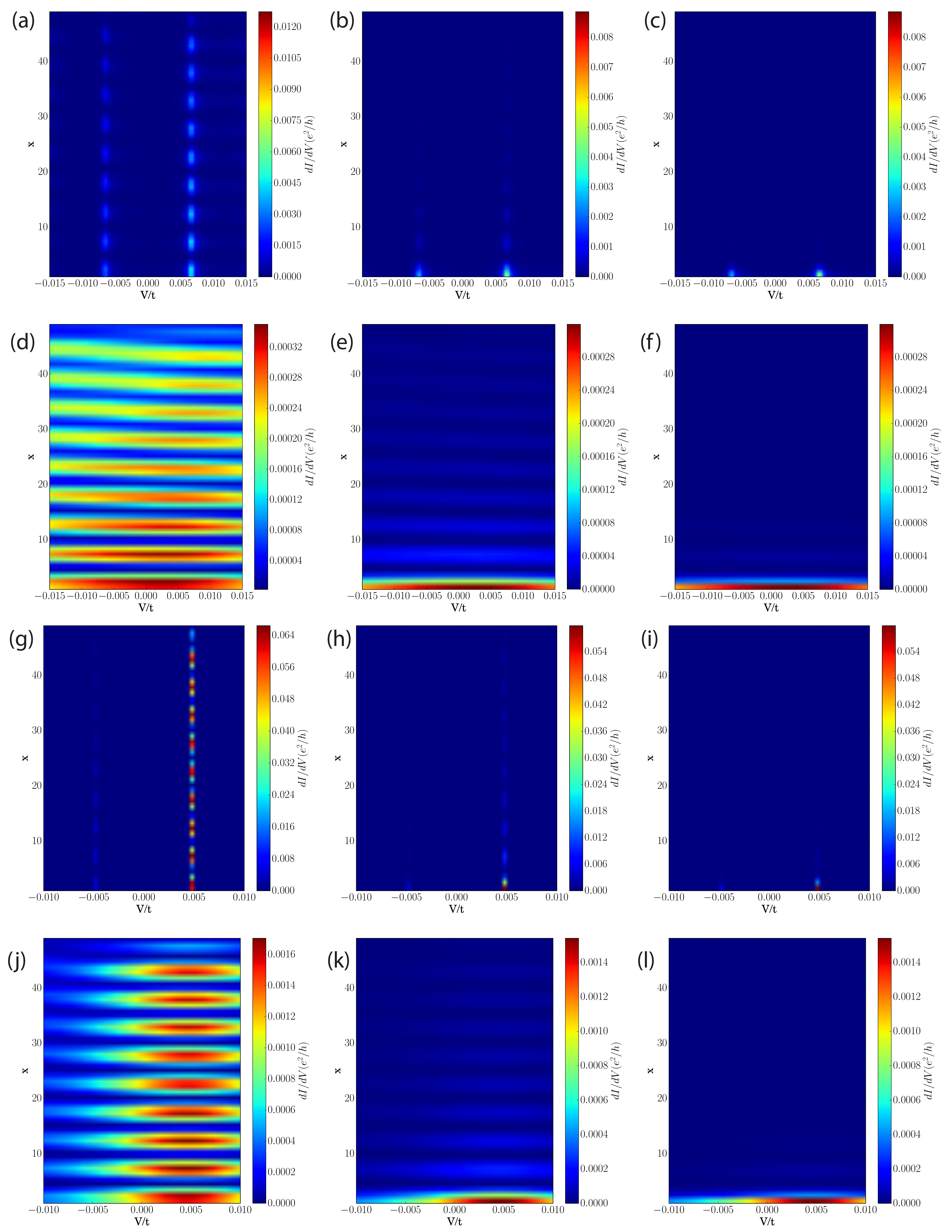} 
\caption{Differential conductance spatial profiles for a system of length  $L=100$ and dimensions $d=1,2,3$ are plotted in the left, middle, and right columns respectively. 
In panels (a-f) we use a pairing potential of $\Delta=0.01t \; (\xi=100)$ while in panels (h-l) use $\Delta=0.005t \; (\xi=200)$.
The zero temperature results are given in panels (a-c, g-i) while we use a finite temperature $T=\Delta$ in panels (d-f, j-l) and in both cases STM - FM coupling is $t'=0.01t$. 
Because the system size is comparable to the coherence length ($L \leq \xi$), the Majorana modes hybridize into conventional quasiparticles, leading to a splitting of the ZBCP as seen in the $T=0$ conductance. 
At finite temperature, this signal is smeared across the entire voltage range of $V=(-1.5 \Delta,1.5 \Delta)$.  
The 2D (middle column) and 3D (right column) data differs from the nanowire data by a pre factor (see text) of $1/x$ and $1/x^2$ respectively, which accounts for rapid decay of the modes into the 2D and 3D bulk respectively.}
\label{fig:finiteT_profile}
\end{figure*}

An additional problem in the interpretation of the experimental data of Ref.~\citenum{Yazdani_14} is the issue of disorder which should very strongly suppress the induced p-wave superconductivity.  Given that the induced gap is $0.1 \; meV$ and the typical electronic energy scales in the ferromagnetic chain (i.e. hopping energy, chemical potential, exchange energy) are all in the eV range, one expects the slightest static fluctuations in the system (e.g. 0.1 \% variation in the locations of the Fe atoms or the presence of any neighboring random impurities near the chain) to completely destroy the topological superconductivity in the system since the $p$-wave pairing is not protected against disorder by Anderson's theorem \cite{Sau_12,Disorder}.
One simple and approximate way to estimate disorder effects here is to ask about the amount of elastic scattering which would be necessary to completely suppress the reported $100 \; \mu eV$ topological gap in Ref.~\citenum{Yazdani_14}.  Equating the reported $p$-wave gap to a disorder induced collisional level broadening of $100 \; \mu eV$ in the Fe chain and using the band parameters estimated in Ref.~\citenum{Yazdani_14} for the system, it is easy to conclude that the electronic mean free path along the Fe chain must be longer than $100 nm$ for the disordered system to manifest any topological gap (assuming the clean system gap to be $100 \; \mu eV$).  This is of course inconsistent with the observation of a topological gap in chains of variable lengths between 5 nm and 15 nm as reported \cite{Yazdani_14} since the wire length serves as a cut off for the maximum possible mean free path in the system.  
One could of course assume that the measured gap already incorporates the disorder effect (starting from a much larger clean topological gap), but this would imply very strong dependence of the measured topological gap on the wire length, not reported in Ref.~\citenum{Yazdani_14}. 

In the next section, we discuss in detail some possibilities for the reconciliation of the experimental observations of Ref.~\citenum{Yazdani_14} with the theoretical Majorana interpretation. 
In particular, several specific ambiguities regarding the Majorana interpretation of the experimental observations described in the current section are shown to become less severe once certain additional elements of physics, which might be playing a role in the measurements of Ref.~\citenum{Yazdani_14}, are taken into account.

\section{Comparison with a recent experiment}

The most compelling qualitative features of the experimental STM results presented in Ref.~\citenum{Yazdani_14}, providing evidence for zero-energy Majorana modes in Fe chains lying on superconducting Pb substrates, are the existence of a (weak and broad) zero-bias differential tunneling conductance peak spatially localized near the ends of the chains which seems to disappear as the STM tip probes the middle regions of the chains away from the ends. 
This observation of a zero-bias peak seemingly spatially localized at the wire ends while being a necessary signature for Majorana modes as pointed out a long time ago\cite{Sengupta-PRB-2001} is, however, not sufficient (see the next section for the chiral fractional Josephson effect which could serve as a sufficient condition). 

The puzzling features of the experimental observation \cite{Yazdani_14} are many: 
(i) the lack of any obvious superconducting gap manifesting in the STM data on the Fe nanowire although a striking gap, in precise quantitative agreement with the BCS theory, shows up for the Pb substrate itself;
(ii) the observed zero-bias peak is minuscule, being reduced by a factor of $10^{-3}$ to $10^{-4}$ from the canonical $2e^2/h$ zero-bias conductance associated with the perfect Andreev reflection from the Majorana mode;
(iii) the zero-bias peak is exceptionally broad, being comparable to the estimated topological superconducting gap in the nanowire; 
(iv) the experimental temperature is comparable to the estimated topological superconducting gap which makes it very difficult, if not impossible, to discuss features associated with Majorana modes; 
(v) the very small topological gap implies very long coherence length, and consequently, long Majorana localization length ($\xi \sim \Delta^{-1}$), which makes the experimental observation of the spatial localization of the Majorana modes to atomic sharpness ($\lesssim0.5 \; nm$) very puzzling to say the least since given that the experimental coherence length is $\xi \sim 100 \; nm$;
(vi) the observed strong particle-hole asymmetry is puzzling since the Majorana mode and the subgap BdG spectrum should obey-particle hole symmetry; 
(vii) given the small topological gap and the consequent long superconducting coherence length in the nanowire, it is unclear why (and how) the Majorana modes could be at zero energy since there should be considerable hybridization between the two end Majorana modes leading to split quasi-Majorana subgap modes away from zero energy;
(viii) connected with the last item, the long coherence length ($\xi >15 \;nm$) and the short lengths of the nanowire ($5-15 \; nm$) used in the experiment imply that considerable Majorana splitting oscillations should manifest themselves in the experiment (since the coherence length is greater than the wire length), which are, however, not seen;
(ix) no obvious Majorana peak (even split peaks) is seen in the very short ($< 5\; nm$) wires;
(x) the STM gap signature for superconductivity which manifests itself strikingly in the Pb substrate itself away from the nanowire, disappears completely on the nanowire without any obvious signature for the Shiba subgap states in the Pb superconducting gap as should be expected for Fe, a magnetic impurity, on Pb, an ordinary $s$-wave superconductor (unless, of course, the weak peaks being identified as Majorana modes are really the subgap Shiba states induced by Fe in Pb).

Some of the problems with the Majorana interpretation of the data in Ref.~\citenum{Yazdani_14} have already been mentioned in the earlier sections of this paper, but we summarize all of them tougher right in the beginning of this section because we will provide detailed numerical results in this section establishing that, in spite of all these problems the observations in Ref.~\citenum{Yazdani_14} are not inconsistent with the Majorana interpretation, but any definitive conclusion would necessitate much lower measurement temperatures and much higher instrumental resolution, as well as a system exhibiting a more definitive topological gap. 

First, the high measurement temperature ($k_B T \gtrsim \Delta_p$) suppresses and broadens all features associated with any induced superconducting gap and all associated subgap features (Majorana or non-Majorana), thus making it difficult to observe any Majorana energy splitting oscillations since the features are simply too weak.
To make this point explicit, we show in Figs.~10-12 our calculated STM tunneling spectra on the superconducting ferromagnetic nanowire at the wire end ($x=0$) for three different temperatures ($T=0,\Delta/2,\Delta$) , two different topological gap values ($\Delta=0.01t, 0.005t$), and four different wire lengths ($L=700,300,200,100$) with a typical coherence lengths being $\xi_1=100$ (for $\Delta=0.01t$) and $\xi_2=200$ (for $\Delta=0.005t$).
We have chosen the temperature, the topological gap, the wire length and the STM tunnel barrier strength at the tip ($t'=0.01t$) to be qualitatively consistent with the experimental results, and, by construction, the voltage bias region ($\pm 1.5 \Delta $) chosen along the abscissa focuses entirely on the possible Majorana physics in the topological gap of the ferromagnetic wire. 
We note that since our model has, by construction, no subgap states other than the Majorana zero modes (which may very well be split due to hybridization in the finite wire), any subgap structure manifesting itself in our results, by definition, arises from the Majorana fermions. 
The most prominent feature of Figs.~10-12 is that, although the $T=0$ Majorana peaks are sharp and may manifest energy splitting (the magnitude of the splitting decreases with increasing wire length) as expected theoretically, the finite temperature differential conductance at $x=0$ for all four wire lengths shows very broad features (with the broadening being comparable to or even larger than the topological gap itself) with a weak smooth peak at zero energy.
For the smallest gap ($\Delta=0.005t$ in Fig.~11) and for the shorter wires ($L=100,200$), the peak is actually asymmetric and in fact goes out of the gap region (see Fig.~11) due to the  Majorana overlap.
These features are all consistent with the experiment where a very weak and very broad zero bias peak is seen only in the longer wires (at the wire ends), and for shorter wires, the peak is not observed.

To understand the absence of any split Majorana peaks or Majorana oscillations away from the ends of the Fe wire in the experiment, we show in Figs.~13-15 the calculated Majorana conductance peaks around the spatial location $x=L/4$ for $L=100,200,300,700$ again for $T=0,\Delta/2,\Delta$ and for $\Delta=0.01t$ and $0.005t$. 
The interesting (and somewhat surprising) point to note is that essentially no Majorana peaks or oscillations are discernible around $x=L/4$ (i.e. Figs.~13-15) for finite temperature although the $T=0$ situation does reflect split peaks. In fact, all traces of Majorana modes, have disappeared completely around $x=L/4$ in the finite temperature results in agreement with the experimental finding, although the superconducting coherence length ($\xi \approx L$, certainly $\xi > L/4$) is long enough that one would expect (as found at $T=0$) signatures of Majorana oscillations at $x=L/4$. 
Again, the high experimental temperature compared with the induced superconducting gap in the nanowire is responsible for the suppression of Majorana oscillations.

Finally in Figs.~16 and 17 we show color plots for the spatial profiles of the differential conductance calculated at $T=0$ (Fig.~16) and $\Delta$ (Fig.~17) respectively, for three different wire lengths ($L=50,100,700$) and one value of the gap ($\Delta=0.01t$).
Results are shown only for finite segments of the wires near the ends, establishing what is already apparent from Figs.~10-15:
(i) the conductance is mostly localized at the wire ends at the finite temperatures of the experiments, and 
(ii) in shorter wires ($L=50$) the Majorana peak is pushed out of the gap due to strong hybridization making it impossible to observe. 
Both of these findings are consistent with experiment as emphasized already. 

We note that there is an additional element of physics, beyond the the scope of the current work, which may lead to further spatial localization of the Majorana modes as observed in the experimental system studied in Ref.~\citenum{Yazdani_14}. 
This is the effective system dimensionally of the experimental setup, which has been assumed to be almost purely one-dimensional in the current work with all substrate degrees of freedom associated with Pb being integrated away. 
Wa have focused on the one-dimensional ferromagnetic nanowire with the only role of the substrate being the inducement of topological superconductivity in the wire through the proximity effect. 
If the full three-dimensional nature of the problem is taken into account, then the Majorana localization (or hybridization) goes effectively as $e^{-r/\xi}/\sqrt{k_F r}$ (2D) or $e^{-r/\xi}/(k_F r)$ (3D) instead of falling off purely exponentially $e^{-r/\xi}$ (1D) \cite{Zyuzin,Li_14,Peng}. 
We note that Ref.~\citenum{Peng} explicitly illustrates how the Majorana decay length exceeds the chain length in the limit of weak hybridization between the host superconductor and magnetic impurities (i.e. the nanowire limit), while if the hybridization is strong (Shiba limit) the modified superconducting band structure indicates a strong renormalization of the Fermi velocity $v_F$ which modifies the Majorana coherence length (recall $\xi \sim v_F / \Delta$).
Obviously a rigorous inclusion of this enhanced localization effect (because of the denominator) is well beyond the scope of the current effective 1D theory, but we could incorporate the denominator approximately by multiplying the Majorana overlap term by $(k_F r)^{-1}$ (2D) or $(k_F r)^{-2}$ (3D) compared with our calculated 1D result. 
It is obvious that such a modification, which effectively takes us from the ferromagnetic nanowire model to the Shiba chain model in an approximate manner, leads to enhanced Majorana mode localization (see Fig.~18) at the wire ends because of the $1/r$ (or $1/r^2$) factor.  On the other hand, this should not modify the temperature dependence (see Fig.~19) as long as the induced topological superconducting gap is comparable to the temperature as appears to be the case in Ref.~\citenum{Yazdani_14}.
We emphasize that the physics of strong Majorana localization discussed in Ref. 60 is qualitatively included in Figs. 18 and 19.

In Figs.~\ref{fig:2D3DSplitting},\ref{fig:finiteT_profile} we show such a comparison among the 1D, 2D, 3D cases which should only be taken in an approximately qualitative sense. 
The Majorana splitting for the isolated 1D system as well as the rescaled splittings in the 2D, 3D cases are illustrated in Fig.~\ref{fig:2D3DSplitting}. 
Since the coherence length for the MBS at each end of the wire is rescaled, the wavefunction overlap in the bulk of the system is also reduced. 
This manifests itself in a reduction of the energy splitting $\Delta E$ between the MBS as seen in Fig.~\ref{fig:2D3DSplitting}. 
In panel (a,b) we have used parameters which correspond to 1D coherence lengths of $\xi=100,200$ sites and have plotted the energy splitting across a range of wire lengths between just  a few atoms and $L=2\xi$. 
The rescaled energy splittings, which are now comparable to the Fe adatom spacing itself (red and green lines in Fig.~\ref{fig:2D3DSplitting}), show a dramatically reduced energy splitting at wire lengths which are actually smaller than the system size. 

The results in Fig.~\ref{fig:finiteT_profile} demonstrate the expected strong spatial Majorana localization near the wire ends in spite of the very small applicable topological gap (and the associated long superconducting coherence length) in the nanowire. 
As already indicated by the results of Figs.~10-17 the very high experimental temperatures (compared with the induced topological superconducting gap) used in Ref.~\citenum{Yazdani_14} make it impossible to observe any Majorana signatures away from the wire ends even in the 1D model, and therefore, the relevance of the higher dimensional models remains unclear at this stage although these higher-dimensional models may explain Majorana localization if it is observed for temperatures much lower than the topological gap, a situation not yet achieved experimentally.
We emphasize that, although the 2D and 3D results at zero temperature in Fig.~\ref{fig:finiteT_profile} look very different from the corresponding 1D results at $T=0$ and manifest strong MBS localization at the wire ends, the finite-temperature results for all three cases look very similar and are qualitatively the same as what is already shown in our Figs.~12-17.  
Thus, unless STM measurements are carried out at temperatures much lower than the topological gap, nothing conclusive can be stated about the experimental situation with the existing data reported in Ref.~\citenum{Yazdani_14}.
The fact that our results in Fig.~19 look very similar to the earlier results presented here based on the pure 1D ferromagnetic nanowire model establishes that at high temperatures, the STM spectra in the two models are similar, and only future low-temperature experiments can distinguish the two situations.

We point out a particular feature of our results which is also apparent in the experimental STM data. 
The sub-gap conductance spectra often manifests strong particle-hole asymmetry. 
This is a special feature of STM spectroscopy \cite{Martin} where some non-equilibrium inelastic effect destroys the expected particle-hole symmetry which is manifestly obvious in the data of Ref.~\citenum{Yazdani_14}.
In our numerical calculation, we simulate this non-equilibrium inelastic scattering process by having an extremely weakly ($\ll t'$) coupled third lead in the system using the Landauer-Buttiker scattering formalism. 

To conclude this section, we have shown by presenting extensive numerical results for the calculated spatially resolved differential conductance spectra for the strongly-coupled ferromagnetic nanowire under proximity-induced $p$-wave superconducting proximity effect that many apparent puzzling features of the experimental data in Ref.~\citenum{Yazdani_14} can be made consistent with the existence of Majorana zero modes in the system once the very high temperature (compared to the superconducting gap) and the very weak tunnel coupling of the STM tip are included in the theory. 
\section{Chiral Fractional Josephson Effect}

\begin{figure}[tb!]
\centering
\includegraphics[width=\columnwidth]{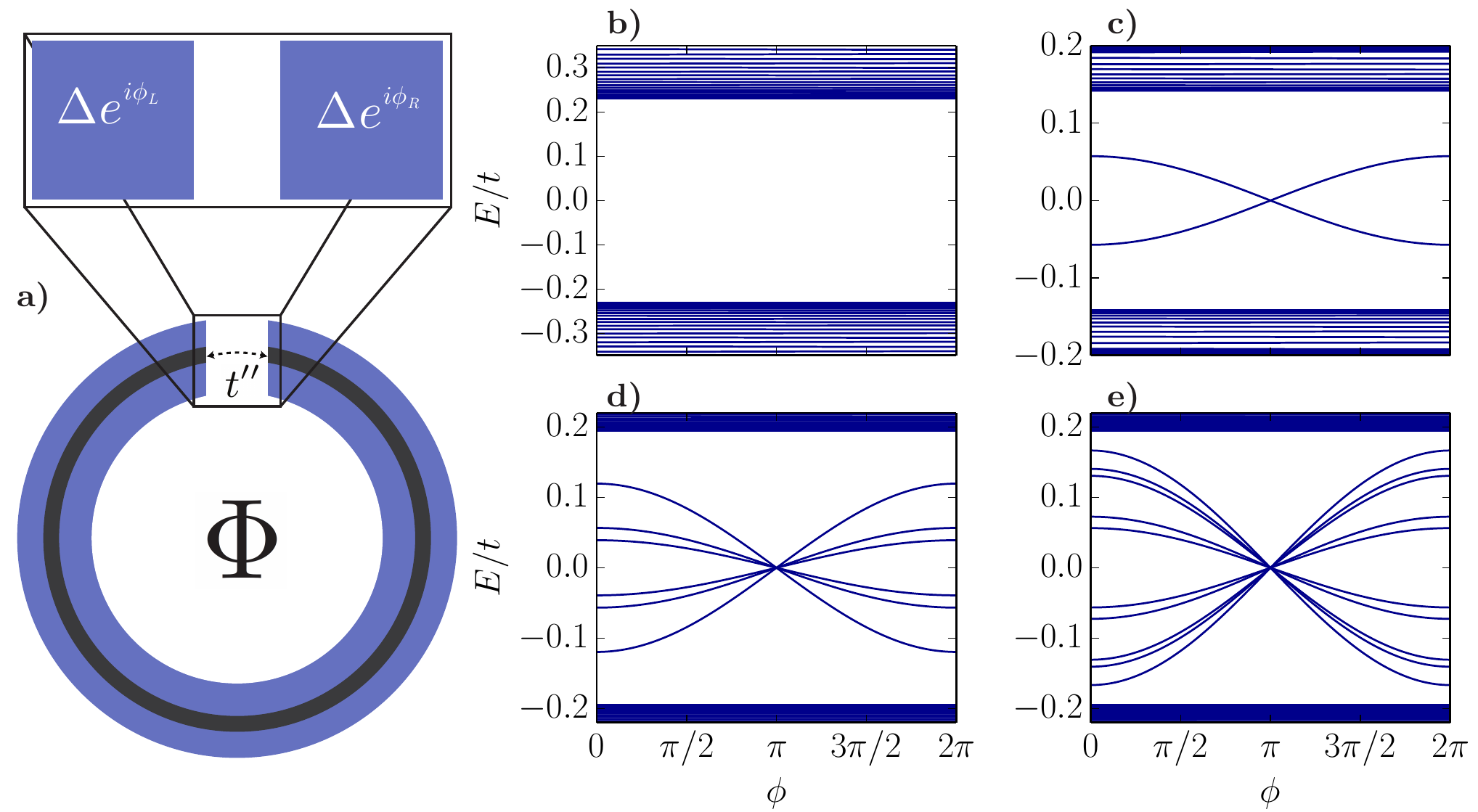}
\caption{a) Top view of ring geometry Josephson junction formed by weakly coupling ($t''=t/10$) the two ends of a nanowire.
The phase difference across the junction is controlled by the magnetic flux, $\Phi$, threading the ring.
b-e) Evolution of subgap Andreev bound states as a function of the phase difference between two superconductors in a Josephson junction setup.
Zero energy crossings are absent in the ABS spectrum in the topologically trivial regime as seen in panel b).
c) In the $|{\cal{W}}|=1$ phase, the ABS spectrum crosses zero energy for the single value $\phi=\pi$.
Transitions between the subgap branches are forbidden by particle-hole symmetry so that Fermion parity flips as $\phi$ evolves by $2\pi$.
d,e) Spectra in a system indexed by topological invariant $|{\cal{W}}|=3,5$.
There are $|{\cal{W}}|$ protected zero energy crossings and chiral symmetry forbids transitions between different sets of subgap branches.}
\label{fig:Josephson}
\end{figure}

We now turn our attention to another important signature which would definitively confirm the presence of the topological superconducting state, the fractional Josephson effect.
A Josephson junction (JJ) consists of two superconductors connected by a weak link, normal metal or insulator for example.
Conventionally, the current passing through a JJ obeys the $2 \pi$ periodic current-phase Josephson relation, $I_J=I_c \sin(\Delta \phi)$, where $\Delta \phi = \phi_L - \phi_R $ is the phase difference across the junction and $I_c$ is the critical supercurrent.
It has been predicted \cite{KSY_04,Kitaev-1D} that spinless p-wave superconductors, when in a topologically non-trivial phase, obey a $4 \pi$ periodic current-phase relation $I_J \propto \sin(\phi/2)$.
Even if quasiparticle relaxation times are shorter than the phase adjustment time, one expects to see a peak in the current noise spectrum at half the Josephson frequency\cite{Houzet_13}.
The difference in periodicity is rooted in the fact that only Cooper pairs can tunnel between conventional superconductors, while single electron tunneling is enabled by MBS, thus doubling the Josephson period from $2\pi$ to $4\pi$ in the process.

Weakly coupling the two ends of a finite wire by a hopping integral $t''$ allows us to model a Josephson junction (see Fig.~\ref{fig:Josephson}).
The Hamiltonian for our ring geometry is $H = H_{TS}+H_J$ where $H_J = \sum_{\sigma} t'' (c^{\dagger}_{N \sigma} c^{  }_{1 \sigma} + \text{H.c.})$.
Hybridizing the two Majorana states by $t''$ results in the formation of Andreev bound states (ABS) in the weak link junction.

Next, we wish to control the superconducting phase difference across our junction by penetrating the ring with a magnetic flux $\Phi$.
Since we introduce a flux we must use the phase which is invariant under a gauge transformations $\bm{A} \mapsto \bm{A} + \nabla \chi$ ($\chi$ is an arbitrary scalar function).
This is given by \cite{Tinkham} $\phi =  \Delta \phi -  \frac{2 \pi}{\Phi_0}\int \bm{A} \cdot d\bm{l}$ where the integration is performed across the junction.
The magnetic vector potential couples to the momentum by $\bm{p} \mapsto \bm{p} - e\bm{A}/c$ and we model this on a lattice with the Peierls substitution $t \mapsto t e^{i e \int \bm{A}.\bm{d} \bm{l} / \hbar }$ where the integral is taken along the two sites which $t$ connects.
The Peierls phase factor is exactly half the gauge invariant phase and the matrix elements connecting the ring junction become $t'' e^{\pm i \phi/2}$ (Fig.~\ref{fig:Josephson}) where the $\pm$ corresponds to counter- or clockwise hopping across the junction.

At zero temperature the Josephson quasiparticle current is $I_J= \sum_n \partial E_n(\phi) / \partial \phi$, where $n$ indicates the nth excited Bogoliubov state \cite{KSY_04} and the summation is taken over all states up to the Fermi energy.
Using this and the Josephson current-phase relation, it is clear that the relevant quasiparticle energies obey a cosine like behavior with the same periodicity as the current.
Andreev bound states (ABS) are quasiparticles bound to the junction and are responsible for the Josephson current, in contrast with $\phi$ independent bulk states which do not contribute to the Josephson current.
Therefore, one can deduce the periodicity of the Josephson current just by looking at the $\phi$ dependence of the ABS spectrum (Fig.~\ref{fig:Josephson}).
A topologically trivial junction is marked by an absence (generically any even number) of zero energy crossings as seen in panel (b).
In this regime each quasiparticle's energy is $2 \pi$ periodic, the fermion parity is conserved, and the Josephson relation is a conventional $2 \pi$ periodic one.
Upon entering the topologically non-trivial regime, a single protected crossing emerges at $\phi = \pi$ (an odd number of crossings will generally flip Fermion parity).
Transitions between the positive and negative energy subgap branches are forbidden by particle-hole symmetry so that Fermion parity flips as $\phi$ evolves by $2\pi$.
During this process, a single electron is transferred across the junction which is forbidden in parity conserving topologically trivial superconductors.
Winding $\phi$ by another factor of $2 \pi$ recovers the spectrum and fermion parity, leading to a $4 \pi$ Josephson effect.
Further increasing the chemical potential causes more energetic sub-bands to become occupied and the system transitions into topological phases characterized by topological index ${\cal{W}}>1$.
The ABS spectrum in this regime contains $|{\cal{W}}|$ sets of protected zero energy crossings at $\phi=\pi$.
In addition to PH symmetry, chiral symmetry now forbids transitions between the different sets of ABS leading to a robust $4 \pi$ Josephson effect.
The basic idea here is that non-trivial JJ physics, indicating the existence of the fractional Josephson effect associated with the existence of the MBS in the system, can be tuned by varying W rather than $\mu$.
$W$ should in principle be easier to tune in the current system, due to the experimental challenges associated with manipulating chemical potential via an electrostatic gate potentials.

Although, as explained in section VI the current experimental results presented in Ref.~\citenum{Yazdani_14} are indeed consistent with the existence of Majorana zero modes in the nanowire, the existing observation of a weak and broad zero bias conductance peak localized at the wire ends at best satisfies the necessary condition for the Majorana modes. 
To make the Majorana existence definitive, the observational of a tunable fractional Josephson effect, as discussed in this section would be necessary. 

\section{Summary and Conclusion}
In this work, motivated by a recent experiment\cite{Yazdani_14} in which STM zero-bias peaks have been observed from the edges of chains of magnetic atoms deposited on superconducting Pb substrates, we show that a multichannel ferromagnetic wire deposited on a spin-orbit coupled superconducting substrate can realize a non-trivial topological superconducting state with one or more Majorana bound states localized at the wire ends. We explain how the persistence of the zero-bias phenomena for generic parameters \cite{Yazdani_14} as observed in the experiments can be understood if the induced topological superconductivity is chiral in nature.
To test this hypothesis, we develop several experimental signatures which may verify the existence of a topological superconducting state.
When several chains of magnetic atoms (such as Fe), or a multichannel FM nanowire, are deposited onto a spin-orbit coupled superconductor such as Pb, a mixture of spin-singlet and triplet components are expected to be proximity induced \cite{Nadj-Perge, Pientka, Brydon, Duckheim, Takei, HBSTS}.
The induced topological superconductivity may generically belong to the topological class BDI, which is characterized by a ${\mathbb{Z}}$ topological invariant.
In this case, by varying the wire width -- or equivalently by coupling parallel chains of magnetic atoms -- one can increase or decrease the strength of the zero-bias signal in a controlled manner.
The width dependence of the peak height is robust against finite temperature effects, which reduce the peak height from the zero-temperature quantized value.
Observing this width dependence of the ZBCP would establish the presence of a chiral topological superconducting state. We also show that the $4\pi$ fractional Josephson
effect remains, even in the presence of multiple spatially
overlapping MBS, and can be used to reveal the Majorana
occupation number by tuning the width parameter of the coupled chains.

This work has been motivated by recent experiments \cite{Yazdani_14} studying the conductance spectrum of a ferromagnetic wire on a spin-orbit coupled superconductor.
Our calculations reproduce the qualitative features of the position and voltage dependence of the conductance experiments.
The low value of the measured conductance in experiment can be attributed to the small topological gap, the high tunnel barrier at the STM-wire contact, and the high temperature.

We show using a superconducting ferromagnetic topological nanowire model, that all the seemingly inconsistent features of the experimental data (e.g. a very weak and very broad zero bias peak arising only from the wire ends, absence of any zero bias peak in the middle of the wire, absence of any observed Majorana splitting oscillations, etc.) can be reconciled once the very high experimental temperature (comparable with the topological superconducting gap) is quantitatively taken into account in the theory. 
Such a high temperature and the extremely small topological gap basically make it impossible to detect any Majorana features in the STM conductance away from the wire ends since all such features are broadened and weakened below the instrumental resolution (which is also comparable to the experimental topological gap).
The experimental regime studied in Ref.~\citenum{Yazdani_14} is highly non-ideal for Majorana investigations since both the experimental temperature and instrumental energy resolution are comparable to the superconducting gap, and any subgap states (e.g. an Andreev bound state or Shiba state or some other fermionic subgap state) would manifest behaviors very similar to the experimental observations \cite{FutureSDS}. 
Basically, the broadening of all subgap peaks would be large enough that nothing conclusive can be stated definitively about their origin at such high temperatures. 
The only way to obtain a more definitive conclusion is to carry out STM spectroscopy at much lower temperatures ($\sim$ one tenth the gap value) so that much sharper zero-bias peaks, as well as Majorana oscillations, are observed as stronger evidence for the existence of the Majorana zero modes. 
Since our model has no subgap states other than the Majorana zero modes by construction, we can conclude that the experimental observations are indeed superficially consistent with the Majorana theory. 
This is, however, an important achievement since at first sight, the very long superconducting coherence length associated with the very small topological gap appears to be prima facie inconsistent with the spatial localization of the Majorana modes at the wire ends. 
We have shown that, given the high temperature and poor experimental resolution (as well as extremely weak SM tunneling strength), that Majorana features away from the wire ends (as well as any Majorana energy splitting effects due to hybridization) are simply not observable in the current experiment. 
We have also shown that these same experimental limitations lead to the non-manifestation of any Majorana-like features in short wires as the split-Majorana zero modes are pushed essentially above the topological gap where thy are unobservable. 
Only going to temperatures much lower than the topological gap values can decisively determine whether the observed weak and broad zero bias peak has anything to do with Majorana bound states or are arising from other subgap states present in the system.

We mention before concluding that very recent theoretical work has appeared in the literature \cite{Li_14,Peng}, after the original submission of our work, suggesting that the ferromagnetic Shiba chain model \cite{Brydon} rather than the ferromagnetic nanowire model \cite{HBSTS} might be the appropriate starting point for the physics underlying the system studied in Ref.~\citenum{Yazdani_14}.  
Notably, Ref.~\citenum{Peng} illustrates that if this indeed the case, the Majorana decay length can be strongly renormalized leading to general qualitative agreement between theory and experiment at zero temperature.  
We have, however, shown that the tunneling conductance at high temperatures (comparable to the induced gap) is qualitatively the same in both the ferromagnetic nanowire (weakly localized Majorana) and the Shiba chain (strongly localized Majorana) model, and thus experiments at temperatures much lower than the induced gap are necessary for any definitive conclusion.  
If the Majorana modes are indeed strongly localized, then the zero-bias peak should become very sharp at low temperatures in long wires whereas clear Majorana splitting should be observable for shorter chains.  Such observations of sharp zero-bias peaks (split peaks) for long (short) chains at low temperature will be the definitive evidence for the experimental observation of topological edge modes for Fe chains on Pb substrates.
However, a very recent work\cite{Sau_Brydon_15} even suggests that the observations of Ref.~\citenum{Yazdani_14} can be simply explained by the appearance of Shiba bound states (rather than Majorana zero modes) in the system without any need to invoke any topological superconductivity.  
We emphasize that ($i$) these theoretical suggestions, while interesting, are not yet experimentally validated, and ($ii$) temperature effects studied in our work are not discussed at all in these works and, based on our results shown in Fig.~19, we conclude that thermal effects are similar in the Shiba chain model and the ferromagnetic nanowire model as long as the induced gap remains comparable to the experimental temperature.  Only temperatures much lower than the induced superconducting gap could distinguish between the Shiba chain and the ferromagnetic nanowire models, and we hope that our results will encourage experiments at much lower temperatures.

We conclude now with a succinct discussion (see the Appendix) of the theoretical novelty of the ferromagnetic nanowire platform as compared with the semiconductor nanowire platform studied extensively in the literature \cite{Long-PRB, Roman, Oreg}. 
One apparent important qualitative difference seems to be that the semiconductor nanowire necessitates a fine-tuning of the Zeeman splitting (or chemical potential) whereas the ferromagnetic platform manifests generic Majorana zero modes without fine-tuning by virtue of its very large exchange-induced spin splitting. 
It turns out, however (the details are provided in the appendix), that the ferromagnetic nanowire is simply a special limiting case of the semiconductor nanowire Majorana problem with $V_z \gg \Delta, \mu$, where $V_z$, $ \Delta$ and $\mu$ are the ferromagnetic exchange spin splitting in the nanowire (or the external magnetic field induced Zeeman spin splitting in the semiconductor nanowire), the superconducting gap and the chemical potential respectively. 
Since the condition $V_z \gg \Delta, \mu$ is presumably satisfied in the experimental system of Ref.~\citenum{Yazdani_14} according to the recent band structure calculations\cite{Li_14}, the ferromagnetic nanowire platform is simply a special (and interesting) limit of the already well-studied semiconductor nanowire case.
We provide these details in the appendix. 

This work is supported by LPS-CMTC and JQI-NSF-PFC at the University of Maryland and by AFOSR (FA9550-13-1-0045) at Clemson University.

\appendix
\section{Connection to the semiconductor nanowire platform}
We show here that the topological properties of the FM wire on an SC substrate in the presence of spin-orbit coupling follow directly from those in the extensively studied  superconductor-semiconductor hybrid system in the limit of very large spin splitting. 
Using Eq.~(33) from Ref.~\citenum{Sau_12} in the semiconductor-superconductor system, we have for the clean-limit quasiparticle gap $\Delta$ in the topological superconducting state:
\begin{equation}
\label{eq:Delta}
\Delta = \left(\frac{ \lambda \Delta_s } {\lambda + \Delta_s} \right)\frac{\alpha k_F}{\sqrt{V_z^2 +\alpha^2 k^2_F}},
\end{equation}
where $\Delta\equiv \Delta_p$ is the proximity induced topological gap (in the Fe wire), $\Delta_s$ is the $s$-wave pairing potential in the substrate (Pb in our case), $\alpha k_F \equiv E_{so}$ is the spin-orbit coupling required to induce $p$-wave proximity pairing from an $s$-wave SC as originally discussed in Ref.~\citenum{Long-PRB}, $V_z \equiv J$ is the (exchange) spin splitting in the Fe nanowire, and $\lambda$ is the tunneling energy coupling the substrate SC (Pb) and the nanowire (Fe), essential for inducing the proximity effect in the nanowire. 
We note that it makes no difference in obtaining Eq.~\ref{eq:Delta} whether the spin-orbit coupling energy $E_{so} (\equiv \alpha k_F)$ arises from the substrate or the nanowire. 
The condition for the existence of the topological phase (and the applicability of Eq.~\ref{eq:Delta} describing the effective $p$-wave topological gap in the nanowire) is the following (c.f. Eq.~(35) in Ref.~\citenum{Sau_12}):
\begin{equation}
\label{eq:topo_condition}
V_z^2 > \Delta^2 +\mu^2 , 
\end{equation}
(i.e. $J^2 > \Delta^2 +\mu^2$) where $\mu$ is the chemical potential in the nanowire. 

Restricting now to the FM nanowire on the SC substrate of our interest here \cite{Yazdani_14}, we consider the extremely large spin-splitting situation given by 
\begin{equation}
\label{eq:FM_limit}
J \equiv V_z \gg \Delta, \mu, E_{so}
\end{equation}
where we conclude that the topological condition is trivially satisfied generically, without tuning of parameters (as long as $J$ is large enough for the Fe system to be a half-metal) as in Eq.~\ref{eq:topo_condition} since spin-splitting is by far the largest energy scale. 
This leads to the immediate (and somewhat trivial) conclusion that, as long as Eq.~\ref{eq:FM_limit} is satisfied, the system is in the topological phase with no fine-tuning necessary.
This is assumed to be the situation in Ref.~\citenum{Yazdani_14} and in Ref.~\citenum{Li_14} (where detailed band structure results for the experimental system are provided) as already emphasized in Ref.~\citenum{HBSTS}

This generic existence of a topological phase in the ferromagnetic nanowire, however, comes with the heavy price of a minuscule topological gap $\Delta_p$ as is obvious by taking the $V_z (\equiv J) \gg \alpha k_F (\equiv E_{so})$ limit in Eq.~\ref{eq:Delta}:
\begin{equation}
\label{eq:minuscule}
\Delta_p \approx \left(\frac{ \lambda \Delta_s } {\lambda + \Delta_s} \right) \left( \frac{E_{so}}{J} \right).
\end{equation}
Note that there are two terms in Eq.~\ref{eq:minuscule} (within the two parenthesis) have different physical origins: the first term, $(\lambda \Delta_s)(\lambda + \Delta_s)^{-1}$, is the proximity effect induced by the substrate itself and the second term, $E_{so}/J$, is the topological effect in the magnetic wire. 
Writing the first term as 
\begin{equation}
\label{eq:Delta_0}
\Delta_0 = (\lambda \Delta_s)(\lambda + \Delta_s)^{-1},
\end{equation}
where $\Delta_0$ is essentially the $s$-wave pairing potential induced in the wire by the substrate we obtain the useful relation: 
\begin{equation}
\label{eq:Delta_p}
\Delta_p \approx \Delta_0 (E_{so}/J).
\end{equation}
We note that Eq.~\ref{eq:Delta_p} has recently been quoted in Ref.~\citenum{Li_14} as a new result without the realization that this formula is simply a special limit of the earlier topological theories studied in the context of semiconductor nanowire Majorana fermions. 
Eq.~\ref{eq:Delta_p} defines the operational equation for the topological induced gap in the ferromagnetic nanowire, establishing clearly that the ferromagnetic wire on a superconducting substrate is not a new Majorana platform at all from the \textit{theoretical perspective}, but is simply a large spin-splitting limit of the well-studied topological semiconductor system. 
This point may not have been emphasized in the recent publications \cite{Yazdani_14,HBSTS,Li_14} on this topic. 

One immediate consequence of the large spin-splitting limit is that that topological gap is suppressed by the factor $J$, which is the largest scale in the problem. 
Since $\Delta \sim  meV$ (at best) and $E_{so} \sim meV$ whereas $J\sim eV$ in the Pb-Fe hybrid system, one gets an agreement with the experimental findings in Ref.~\citenum{Yazdani_14}
\begin{equation}
\label{eq:Delta_p_estimate}
\Delta_p \sim 100 \; \mu eV \; \text{(at best)}
\end{equation}
Thus, a minuscule topological gap is the price one must pay to have a generic topological phase in the ferromagnetic nanowire problem. 
We see no escape from this small gap conundrum for the Fe on Pb topological platform. 

Using the semiconductor nanowire analogy, we can also estimate the effect of disorder on the topological gap. 
It is known that although the disorder in the substrate superconductor has no effect on the topological gap (by virtue of Anderson's theorem) any disorder in the Fe wire itself has a profound detrimental effect \cite{Sau_12,Disorder} on the topological gap since p-wave superconductivity is vulnerable to elastic scattering as Anderson's theorem does not apply in the topological wire itself. 

Using the existing work in semiconductor nanowires \cite{Sau_12,Disorder} we have (see Eq.(20) in Ref.~\citenum{Sau_12}) the following condition for the requisite elastic disorder scattering time for topological superconductivity to be possible in the hybrid system:
\begin{equation}
\tau \gg J/(\Delta_s E_{so}).
\end{equation}
This indicates that the system is highly susceptible to disorder in the large $J$ limit (since $J\gg \Delta_s, E_{so}$, we conclude that we must have $\tau \Delta_s \gg1$,
requiring very long scattering times for the survival of topological superconductors) and we can easily obtain a crude numerical estimate by equating the induced topological gap $\Delta_p$ with the disorder-induced level broadening $\hbar/2\tau$ to write the condition for the survival of the topological superconductivity in the Fe wire as:
\begin{equation}
\tau \gg \frac{2 \Delta_p}{\hbar}.
\end{equation}
Converting the above condition to a minimum necessary mean free path ($l$) in the Fe chain, by using $\Delta_p \approx 100 \; \mu eV$ as quoted in Ref.~\citenum{Yazdani_14} and from the known material parameters for the Fe wire, we conclude that the following condition on the Fe chain mean free path would be necessary for the observation of the Majorana zero modes: 
\begin{equation}
l > 500 \; nm.
\end{equation}
Since the nanowires used in Ref.~\citenum{Yazdani_14} are shorter that this minimum required mean free path (and by definition, a mean free path cannot be longer than the system length), we conclude that any Majorana zero mode would be strongly suppressed by disorder effects in the Fe nanowire. 

A simple way of appreciating the disorder problem in the Fe/Pb system is to note that the reported topological gap ($\sim 100 \; \mu eV$) is $\sim 10^{-4}t$, where $t$ is the typical hopping energy in the Fe chain. 
This means that any fluctuation or error in $t$ arising, for example, from the expected lack of an absolutely precise periodic arrangement of the Fe atoms along the chain (or from environmental impurities and defects) must be well below 1 part in $10^4$ for a robust topological gap. 
This is a rather demanding requirement.


\begin{thebibliography}{99}
\bibitem{Majorana} E. Majorana, Nuovo Cimento \textbf{14}, 171 (1937).

\bibitem{Wilczek} F. Wilczek, Nature Physics \textbf{5}, 614 (2009).

\bibitem{Read-Green} N. Read and D. Green, Phys. Rev. B \textbf{61}, 10267 (2000).

\bibitem{Kitaev-1D} A. Y. Kitaev, Physics-Uspekhi \textbf{44}, 131 (2001).

\bibitem{Ivanov} D. A. Ivanov, Phys. Rev. Lett. {\bf 86}, 268 (2001).

\bibitem{Nayak_2008} C. Nayak, S. H. Simon, A. Stern, M. Freedman, S. Das Sarma, Rev. Mod. Phys. {\bf 80}, 1083 (2008).

\bibitem{Tewari-strontium} S. Das Sarma, C. Nayak, S. Tewari, Phys. Rev. B \textbf{73}, 220502 (R) (2006)

\bibitem{Fu-Kane}L. Fu and C. L. Kane, Phys. Rev. Lett. \textbf{100}, 096407 (2008)

\bibitem{Zhang-Tewari}C. W. Zhang, S. Tewari, R. M. Lutchyn, S. Das Sarma, Phys. Rev. Lett. \textbf{101}, 160401 (2008);
S. Tewari, S. Das Sarma, C. Nayak, C. Zhang, and P. Zoller, PRL {\bf 98}, 010506 (2007).

\bibitem{Sato-Fujimoto}M. Sato, Y. Takahashi, S. Fujimoto, Phys. Rev. Lett. \textbf{103}, 020401 (2009).

\bibitem{Sau-Generic} Jay D. Sau, R. M. Lutchyn, S. Tewari, S. Das Sarma, Phys. Rev. Lett. \textbf{104}, 040502 (2010).

\bibitem{Tewari-Annals} S. Tewari, J. D. Sau, S. Das Sarma,  Ann. Phys. \textbf{325}, 219 (2010).

\bibitem{Long-PRB} J. D. Sau, S. Tewari, R. Lutchyn, T. Stanescu and S. Das Sarma, Phys. Rev. B \textbf{82}, 214509 (2010).

\bibitem{Roman} R. M. Lutchyn, J. D. Sau, S. Das Sarma, Phys. Rev. Lett. \textbf{105}, 077001 (2010).

\bibitem{Oreg} Y. Oreg, G. Refael, F. von Oppen, Phys. Rev. Lett. \textbf{105}, 177002 (2010).


\bibitem{Mourik} V. Mourik, K. Zuo,  S. M. Frolov, S. R. Plissard, E. P. A. M. Bakkers and L. P. Kouwenhoven, Science {\bf 336}, 1003 (2012).

\bibitem{Deng} M. T. Deng, C. L. Yu, G. Y. Huang, M. Larsson, P. Caroff, H. Q. Xu,  Nano Lett. {\bf 12},  6414 (2012).

\bibitem{Weizman}A. Das, Y. Ronen, Y. Most, Y. Oreg, M. Heiblum, H. Shtrikman, Nature Physics {\bf 8}, 887 (2012).

\bibitem{Rokhinson} L. P. Rokhinson, X. Liu, J. K. Furdyna, Nature Physics {\bf 8}, 795 (2012).

\bibitem{Churchill} H. O. H. Churchill, V. Fatemi, K. Grove-Rasmussen, M. T. Deng, P. Caroff, H. Q. Xu, C. M. Marcus, Phys. Rev. B \textbf{87}, 241401(R) (2013)

\bibitem{Finck} A. D. K. Finck, D. J. Van Harlingen, P. K. Mohseni, K. Jung, X. Li, Phys. Rev. Lett. \textbf{110}, 126406 (2013).


\bibitem{Stanescu} T. D. Stanescu, S. Tewari, J. Phys. Condens. Matter \textbf{25}, 233201 (2013)


\bibitem{AZ} A. Altland and M. R. Zirnbauer, Phys. Rev. B \textbf{55}, 1142 (1997)

\bibitem{Schnyder_2008}A. P. Schnyder, S. Ryu, A. Furusaki, and A. W. W. Ludwig, Phys. Rev. B \textbf{78} 195125 (2008);
A. P. Schnyder, S. Ryu, A. Furusaki, and A. W. W. Ludwig,  AIP Conf. Proc. \textbf{1134} 10 (2009).

\bibitem{Kitaev_2009} A. Yu Kitaev AIP Conf. Proc. \textbf{1134} 22 (2009).

\bibitem{Ryu_2010} S. Ryu, A. Schnyder, A. Furusaki, A. W. W. Ludwig, New J. Phys. \textbf{12}, 065010 (2010)


\bibitem{Sato_11}M. Sato, Y. Tanaka, K. Yada, and T. Yokoyama, Phys. Rev. B \textbf{83}, 224511 (2011)

\bibitem{Law_12} C. L. M. Wong, K. T. Law, Phys. Rev. B \textbf{86}, 184516 (2012)

\bibitem{Nakosai_13} S. Nakosai, J. C. Budich, Y. Tanaka, B. Trauzettel, and N. Nagaosa, PRL \textbf{110}, 117002 (2013).

\bibitem{Deng_12} S. Deng, L. Viola, and G. Ortiz, Phys. Rev. Lett. \textbf{108}, 036803 (2012)

\bibitem{Keselman_13} A. Keselman, L. Fu, A. Stern, and E. Berg, Phys. Rev. Lett. \textbf{111}, 116402 (2013)

\bibitem{Zhang_13} F. Zhang, C. L. Kane, and E. J. Mele, Phys. Rev. Lett. \textbf{111}, 056402 (2013)

\bibitem{Law_14} X. J. Liu, C. L. M. Wong, K. T. Law, Phys. Rev. X \textbf{4}, 021018 (2014)

\bibitem{Deng_13} S. Deng, G. Ortiz, L. Viola, Phys. Rev. B \textbf{87}, 205414 (2013)

\bibitem{Flensberg_13} E. Gaidamauskas, J. Paaske, K. Flensberg, PRL \textbf{112}, 126402 (2014)

\bibitem{Dumitrescu_TR} E. Dumitrescu, J. D. Sau, S. Tewari, arXiv:1310.7938 (2013).

\bibitem{TS_BDI} S. Tewari and J. D. Sau, Phys. Rev. Lett. \textbf{109}, 150408 (2012).

\bibitem{minigap} S. Tewari, T. D. Stanescu, J. D. Sau, and S. Das Sarma, Phys. Rev. B {\bf 86}, 024504 (2012).

\bibitem{Chakravarty} Y. Niu, S. B. Chung, C.-H. Hsu, I. Mandal, S. Raghu, and S. Chakravarty, Phys. Rev. B {\bf 85}, 035110 (2012).

\bibitem{Diez} M. Diez, J. P. Dahlhaus, M. Wimmer, and C. W. J. Beenakker, Phys. Rev. B \textbf{86}, 094501 (2012)

\bibitem{Wong_13} C. L. M. Wong, J. Liu, K. T. Law, P. A. Lee, Phys. Rev. B {\bf 88}, 060504(R) (2013)

\bibitem{He} J. J. He, J. Wu, T.-P. Choy, X.-J. Liu, Y. Tanaka, and K. T. Law, Nat. Commun. {\bf 5}, 3232 (2014).

\bibitem{Dumitrescu} E. Dumitrescu and S. Tewari, Phys. Rev. B \textbf{88}, 220505(R) (2013)

\bibitem{Yazdani_14} S. Nadj-Perge, I. K. Drozdov, J. Li, H. Chen, S. Jeon, J. Seo, A. H. MacDonald, B. A. Bernevig, A. Yazdani, Science {\bf 346}, 602 (2014)

\bibitem{Lee_09} P.A. Lee, arXiv:0907.2681(2009).

\bibitem{Duckheim} M. Duckheim and P. W. Brouwer, Phys. Rev. B {\bf 83}, 054513 (2011);

\bibitem{Chung_11} S.-B. Chung, H.-J. Zhang, X.-L. Qi, and S.-C. Zhang, Phys. Rev. B {\bf 84}, 060510(R)  (2011);

\bibitem{Takei} S. Takei and V. Galitski, Phys. Rev. B {\bf 86}, 054521 (2012)

\bibitem{Samarth} J. Wang et al. Nat. Phys. 6, 389 (2010)

\bibitem{Brydon}  P. M. R. Brydon, H.-Y. Hui, J. D. Sau, arXiv:1407.6345 (2014).

\bibitem{HBSTS} H.-Y. Hui, P. M. R. Brydon, J. D. Sau, S. Tewari and S. Das Sarma arXiv:1407.7519 (2014).

\bibitem{Nadj-Perge} S. Nadj-Perge, I. K. Drozdov, B. A. Bernevig, and A. Yazdani, Phys. Rev. B {\bf 88}, 020407(R) (2013).

\bibitem{Pientka} F. Pientka, L. I. Glazman, and F. von Oppen, Phys. Rev. B {\bf 88}, 155420 (2013).

\bibitem{Glazman} F. Pientka, L. I. Glazman, and F. von Oppen,  Phys. Rev. B {\bf 89}, 180505(R) (2014).

\bibitem{Choy} T.-P. Choy, J. M. Edge, A. R. Akhmerov, and C. W. J. Beenakker, Phys. Rev. B {\bf 84}, 195442 (2011).

\bibitem{YuShibaRusinov} L. Yu, Acta Phys. Sin. 21, 75 (1965); H. Shiba, Prog. Theor. Phys. 40, 435 (1968);
A. I. Rusinov, Zh. Eksp. Teor. Fiz. Pisma. Red. 9, 146 (1968) [JETP Lett.9, 85 (1969)]

\bibitem{Loss} J. Klinovaja, P. Stano, A. Yazdani, and D. Loss, Phys. Rev. Lett. {\bf 111}, 186805 (2013).

\bibitem{Franz} M. M. Vazifeh and M. Franz, Phys. Rev. Lett. {\bf 111}, 206802 (2013)

\bibitem{Lutchyn} Y. Kim, M. Cheng, B. Bauer, R. M. Lutchyn, and S. Das Sarma, Phys. Rev. B {\bf 90}, 060401(R) (2014)

\bibitem{Peng} Y. Peng, F. Pientka, L. I. Glazman, F. von Oppen, arXiv:1412.0151 (2014)

\bibitem{Sau_Robustness} J.D. Sau, R. M. Lutchyn, S. Tewari, S. Das Sarma, Phys. Rev. B 82, 094522 (2010)

\bibitem{Stanescu_Proximity} T. D. Stanescu, J. D. Sau, R. M. Lutchyn, S. Das Sarma, Phys. Rev. B 81, 241310 (2010)

\bibitem{Sengupta-PRB-2001} K. Sengupta, Igor Zutic, Hyok-Jon Kwon, Victor M. Yakovenko, S. Das Sarma,
Phys. Rev. B \textbf{63}, 144531 (2001)

\bibitem{Hasan-Kane} M. Z. Hasan and C. L. Kane, Rev. Mod. Phys. \textbf{82}, 3045 (2010)

\bibitem{Budich-Ardonne2} J.C. Budich and E. Ardonne, Phys. Rev. B \textbf{88}, 134523 (2013)

\bibitem{Qu} C. Qu, M. Gong, Y. Xu1, S. Tewari, and C. Zhang, arXiv:1310.7557 (2013);
E. Dumitrescu, T. D. Stanescu, S. Tewari arXiv:1403.3093 (2014)

\bibitem{BTK} G. E. Blonder, M. Tinkham, and T. M. Klapwijk, Phys. Rev. B \textbf{25}, 4515 (1982)

\bibitem{San-Jose} P. San-Jose, E. Prada, and R. Aguado, PRL \textbf{108}, 257001 (2012)

\bibitem{Kwant} C. W. Groth, M. Wimmer, A. R. Akhmerov, X. Waintal, arXiv:1309.2926 (2013).

\bibitem{Flensberg} K. Flensberg, Phys. Rev. B {\bf 82}, 180516(R) (2010).

\bibitem{Cheng} M. Cheng, R. M. Lutchyn, V. Galitski, S. Das Sarma, Phys. Rev. Lett. {\bf 103}, 107001 (2009);
M. Cheng, R. M. Lutchyn, V. Galitski, S. Das Sarma, Phys. Rev. B {\bf 82}, 094504 (2010); 
S. Das Sarma, J. D. Sau, T. D. Stanescu, Phys. Rev. B {\bf 86}, 220506 (2012);
D. Rainis,  L. Trifunovic, J. Klinovaja, and D. Loss Phys. Rev. B {\bf 87}, 024515 (2013).

\bibitem{Sau_12} J. D. Sau, S. Tewari, S. Das Sarma, Phys. Rev. B {\bf 85}, 064512 (2012);

\bibitem{Disorder}  
J. D. Sau, S. Das Sarma, Phys. Rev. B {\bf 88}, 064506 (2013);
P. W. Brouwer, M. Duckheim, A. Romito, F. von Oppen, Phys. Rev. Lett. {\bf 107}, 196804 (2011);
P. W. Brouwer, M. Duckheim, A. Romito, F. von Oppen, Phys. Rev. B {\bf84},144526 (2011); 
A. M. Lobos, R. M. Lutchyn, S. Das Sarma, Phys. Rev. Lett. {\bf109}, 146403 (2012); 
T. D. Stanescu, R. M. Lutchyn, S. Das Sarma, Phys. Rev. B {\bf84}, 144522 (2011); 
J. Liu, A. C. Potter, K.T. Law, P. A. Lee, Phys. Rev. Lett. {\bf109}, 267002 (2012); 
D. Bagrets and A. Altland Phys. Rev. Lett. {\bf 109}, 227005 (2012);
D. I. Pikulin, J. P. Dahlhaus, M. Wimmer, H. Schomerus, C. W. J. Beenakker, New J. Phys. {\bf 14}, 125011 (2012); 
E. Prada, P. San-Jose, R. Aguado, Phys. Rev. B {\bf 86}, 180503(R) (2012); 
C.-H. Lin, J. D.  Sau, S. Das Sarma, Phys. Rev. B {\bf 86}, 224511(2012); 
I. C. Fulga, F. Hassler, A. R. Akhmerov, C. W. J. Beenakker Phys. Rev. B {\bf 83} 155429 (2011); 
R. M. Lutchyn, T. D. Stanescu, S. Das Sarma, Phys. Rev. B {\bf 85}, 140513(R) (2012).

\bibitem{Zyuzin} A. A. Zyuzin, D. Rainis, J. Klinovaja, and D. Loss, Phys. Rev. Lett. {\bf 111}, 056802 (2013).

\bibitem{Li_14}	J. Li, H. Chen, I. K. Drozdov, A. Yazdani, B. A. Bernevig, A. H. MacDonald, arXiv:1410.3453 (2014); 

\bibitem{Martin} I. Martin and A. F. Morpurgo, Phys. Rev. B {\bf 85}, 144505 (2012);

\bibitem{KSY_04} H.-J. Kwon, K. Sengupta, Victor M. Yakovenko, The European Physical Journal B {\bf 37}, 349(2004);
H.-J. Kwon, V. M. Yakovenko, and K. Sengupta, Low Temp. Phys. {\bf 30}, 613 (2004)

\bibitem{Houzet_13} M. Houzet, J. S. Meyer, D. M. Badiane and L. I. Glazman, PRL \textbf{111}, 046401 (2013);

\bibitem{Tinkham} M. Tinkham, \textit{Introduction to Superconductivity}, 2d Ed. McGraw-Hill, New York, (1996);

\bibitem{FutureSDS} S. Das Sarma, H.-Y. Hui, P.M.R. Brydon, J.D. Sau, to be published

\bibitem{Sau_Brydon_15} J.D. Sau and P.M.R. Brydon, arXiv:1501.03149 (2015)


\end{thebibliography}
\end{document}